\documentclass[10pt,journal,twoside,compsoc]{IEEEtran}

\usepackage{booktabs} 
\usepackage{graphicx}
\usepackage{subfigure}
\usepackage{amsfonts}
\usepackage{amssymb}
\usepackage{color}
\usepackage{epstopdf}
\usepackage{amsmath}
\usepackage{balance}
\ifCLASSOPTIONcompsoc
  \usepackage[nocompress]{cite}
\else
  \usepackage{cite}
\fi

\ifCLASSINFOpdf
\else
\fi
\begin{document}

\title{Securing IoT Devices by Exploiting Backscatter Propagation Signatures}

\author{Zhiqing Luo,
        Wei Wang,~\IEEEmembership{Senior Member,~IEEE,}
        Qianyi Huang,~\IEEEmembership{Member,~IEEE,}
        Tao Jiang,~\IEEEmembership{Fellow,~IEEE,}
        and Qian Zhang,~\IEEEmembership{Fellow,~IEEE}
\IEEEcompsocitemizethanks{\IEEEcompsocthanksitem Part of this work has been presented at ACM SenSys 2018~\cite{luo2018shieldscatter}.

\IEEEcompsocthanksitem This work was supported in part by the National Key R\&D Program of China under Grant 2020YFB1806600, National Science Foundation of China with Grant 62071194, 91738202, 62002150, RGC under Contract CERG 16204418, Contract 16203719, Contract 16204820, Contract R8015, and Guangdong Natural Science Foundation under Grant 2017A030312008. (Corresponding author: Wei Wang.)

\IEEEcompsocthanksitem Z. Luo, W. Wang and T. Jiang are with the School of Electronic Information and Communications, Huazhong University of Science and Technology, Wuhan, Hubei, China.\protect\\
E-mail: \{zhiqing\_luo, weiwangw, taojiang\}@hust.edu.cn.
\IEEEcompsocthanksitem Q. Huang is with the Institute of Future Networks, Southern University of Science and Technology and Peng Cheng Laboratory, China.\protect\\
E-mail:   huangqy@sustech.edu.cn.
\IEEEcompsocthanksitem  Q. Zhang is with the Department of Computer Science and Engineering, Hong Kong University of Science and Technology, Hong Kong, China.\protect\\
E-mail:  qianzh@cse.ust.hk.
}
}

%
%

\markboth{IEEE TRANSACTIONS ON MOBILE COMPUTING,~Vol.~00, No.~00, 00~0000}%
{Luo ET AL.: Securing IoT Devices by Exploiting Backscatter Propagation Signatures}
%




\sloppy
\IEEEcompsoctitleabstractindextext{
\begin{abstract}
The low-power radio technologies open up many opportunities to facilitate Internet-of-Things (IoT) into our daily life, while their minimalist design also makes IoT devices vulnerable to many active attacks. Recent advances use an antenna array to extract fine-grained physical-layer signatures to identify the attackers, which adds burdens in terms of energy and hardware cost to IoT devices. In this paper, we present ShieldScatter, a lightweight system that attaches low-cost tags to single-antenna devices to shield the system from active attacks. The key insight of ShieldScatter is to intentionally create multi-path propagation signatures with the careful deployment of tags. These signatures can be used to construct a sensitive profile to identify the location of the signals’ arrival, and thus detect the threat. In addition, we also design a tag-random scheme and a multiple receivers combination approach to detect a powerful attacker who has the strong priori knowledge of the legitimate user. We prototype ShieldScatter with USRPs and tags to evaluate our system in various environments. The results show that even when the powerful attacker is close to the legitimate device, ShieldScatter can mitigate 95\% of attack attempts while triggering false alarms on just 7\% of legitimate traffic.
\end{abstract}

}

\maketitle

\IEEEdisplaynontitleabstractindextext


%
\IEEEpeerreviewmaketitle

\IEEEraisesectionheading{\section{Introduction}\label{sec:introduction}}

\begin{figure}[t]
	\center
	\includegraphics[width=1.7in]{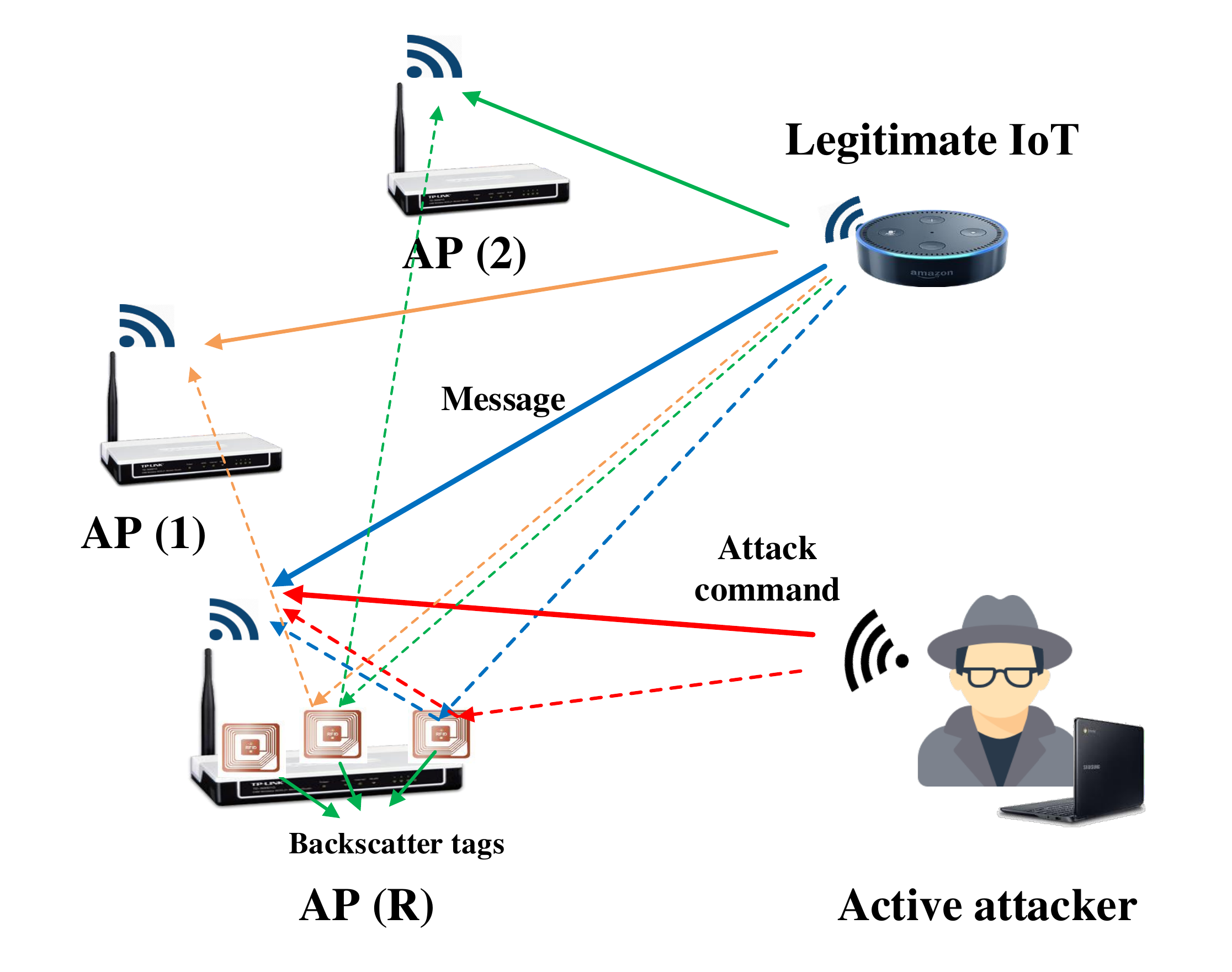} 
	\caption{Illustration of active attack.}
	\label{fig:sketch}\vspace{-0.3cm}
\end{figure}

%
%
%
%
\IEEEPARstart{T}{he} continuous advancement in low power radios and lightweight protocols are driving the proliferation of Internet-of-Things (IoT) in our daily life. However, the minimalist design of IoT devices is like the two sides of a coin: it enables IoT devices to communicate with low power and hardware cost while making IoT devices easily bear the risks of active attacks during devices pairing or data transmission. Many efforts have shown that wireless connectivity can be easily compromised with recent security protocols like WPA2, where an attacker can send unauthorized commands to attack IoT devices, such as spoofing attack and Denial-of-Service (DoS) attack~\cite{8302842,bertka2012802,tews2009practical}. For example, as shown in Fig.~\ref{fig:sketch}, a legitimate IoT device is pairing or sharing data with an access point, that is AP (R). An active attacker equipped with an omnidirectional or directional antenna impersonates the legitimate IoT device and sends a fake command (e.g., DoS command or fake data) to spoof the AP, from which she can successfully eavesdrop users' traffic, mislead the users, or modify the configuration of network~\cite{mantas2011security}.

Many attempts and efforts to defend against attackers for wireless devices have been extensively studied during the past several years. Traditional approaches mainly rely on the complex encryption algorithm, which leads to computational resource and energy wastage~\cite{gehrmann2004manual,dietrich2007financial} and are not feasible for IoT devices that lack sophisticated user interface. Alternatively, fine-grained physical-layer signatures, such as angle of arrival~(AoA)~\cite{xiong2013securearray}, channel state information (CSI)~\cite{jiang2013rejecting} and received signal strength (RSS)~\cite{cai2011good} have recently received much attention to mitigating these threats. However, these systems require at least two antennas or a large antenna array (e.g., an eight-antenna array) to construct sensitive signatures, and thus are not applicable for the systems where the APs and IoT devices are equipped with only a small number of antennas. In addition, location distinction by exploiting the spatial uncorrelation property of wireless channels has been explored for many years~\cite{xiao2009channel}. However, it has been proved that a powerful attacker can create any wireless channel characteristic to deteriorate the location distinction capability of the receiver~\cite{fang2014you}.

To overcome these predicaments, this paper presents ShieldScatter, a lightweight system to secure IoT device pairing and data transmission. Instead of relying on a large antenna array, ShieldScatter advocates the use of merely several ultra-low-cost backscatter tags~\cite{liu2013ambient} to secure the IoT device. Our key insight is that backscatter tags, communicating based on reflecting ambient signals, can be exploited to intentionally create fine-grained multipath signatures. In particular, upon detecting any suspicious transmission, the legitimate device is asked to transmit the challenge-response based signals to the AP within the coherence time. At the same time, the AP controls the tags with a well-designed tag-random scheme to reflect the signals. We observe that, even though the IoT device is attacked by a powerful attacker who can create wireless channel signature to spoof the AP, the propagation signatures created by the backscatter tags can be used to construct a unique profile to identify the device. This unique profile then confirms whether these two signals are from the legitimate IoT device and help defend against active attacks. Moreover, when there are multiple APs, we also combine all of them to construct a more stable profile to detect the attacker.

To realize the above idea, we entail the following challenges.

\textit{(1) How to employ backscatter tags to create sensitive propagation signatures without using an expensive antenna array or other powerful hardware?}
Most home devices employ only a small number of antennas for data transmission, which makes it inapplicable for them to construct accurate propagation signatures. To address this problem, we attach the tags around the AP. When initializing the system, the AP receives the signals. At the same time, the AP controls the tags to reflect wireless signals in turn. This intentional deployment can create artificial multi-path propagations that are sensitive to the senders' location. If these two signals are from the same devices, the multi-path effects of the backscatter tags on these two signals will have a strong similarity when they are within coherence time. Thus, these distinctive propagation signatures can be used to identify a legitimate IoT device.

\textit{(2) How to construct reliable signatures when unstable factors exist?}
In home environments, people walking around, environmental noise and an imperfect circuit design of the tags would lower the similarity of the signals from the devices. In order to construct reliable signatures, ShieldScatter extracts the representative features from the signals for the first step. Then, ShieldScatter aligns and compares the similarity of the features using dynamic time warping~(DTW). Furthermore, a one-class support vector machine~(SVM) classifier is used to distinguish and defend against signals from active attackers. If the signals are from the same device, it will lead to strong similarity and short DTW distances, and then these signals will be clustered into the legitimate class. Otherwise, it will lead to large DTW distance and ShieldScatter can easily defend against the active attacker.

\textit{(3) How to defend against a powerful attacker who has the priori knowledge of all the devices?}
In some case, the IoT device may be attacked by a powerful attacker that has the priori knowledge of all the devices, including the carrier frequency, the  modulation schemes, and so on. With the knowledge, even though we have created the multi-path propagation signatures with tags, this attacker can also impersonate the legitimate IoT device by generating similar wireless signatures to spoof the AP~\cite{fang2014you}. To overcome this predicament, we design a tag-random scheme to defend against this powerful attacker. In particular, the tag-random approach takes into consideration of randomizing the real-time channel states, making the powerful attacker fail to estimate the real multipath channel. Besides,  when there are multiple APs in a room at the same time, the signals reflected by the backscatter tags can also be received at all these receivers. Then, the APs can extract the features from the received signals and combine them to construct a more stable profile, and thus we can improve the accuracy.

\textbf{Summary of result.}
We prototype ShieldScatter with USRPs and ambient backscatter tags to evaluate our system in various environments. The experimental results show that even when the powerful attacker is close to the legitimate device, ShieldScatter can mitigate 95\% of attack attempts while at the same time triggering false alarms on just 7\% of legitimate traffic.

\textbf{Contributions.}
First, we propose ShieldScatter to use the multi-path propagation signatures intentionally created by backscatter tags to secure the IoT device, which avoids the employment of a large antenna array to obtain fine-grained signatures. Second,  we present a tag-random scheme and a multiple receivers combination approach, which successfully defend against a powerful attacker who has the strong priori knowledge of the devices. Finally, we implement a prototype of ShieldScatter and evaluate the performance our system. The results show that our system is robust even when an attacker is close to the legitimate IoT device.

\section{Motivation}
In this section, we first discuss the potential threats to the IoT devices and argue that a lightweight mechanism designed for securing these devices is critical. Next, we investigate the fine-grained signatures of radio propagations used to defend against active attacks, which motivates our design of using backscatter tags to secure IoT devices.

\begin{figure}[t]
	\center
	\includegraphics[width=1.6in]{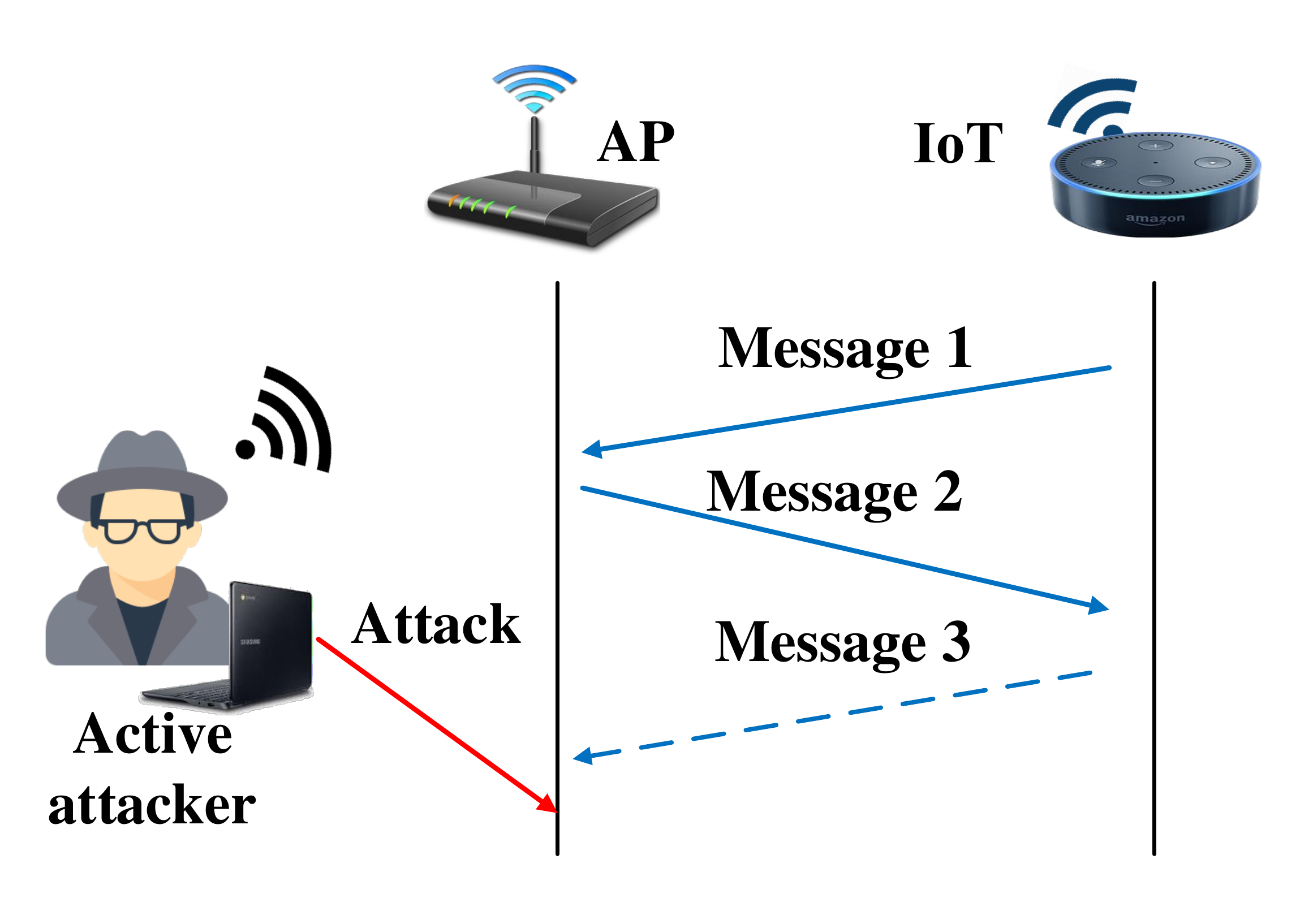} 
	\caption{The active attacker sends a DoS command or fake data to the AP when the IoT device is pairing or exchanging data with the AP. }
	\label{fig:attack_model}\vspace{-0.3cm}
\end{figure}
\subsection{Threat Model}\label{sec:threat}
Recently, the link establishment and data sharing IoT devices and AP have become more and more indispensable. In a smart home, an IoT device needs to establish the connection with an AP when they are trying to share message with each other. However, this establishment process is easily attacked by an active attacker. For example, as shown in Fig.~\ref{fig:attack_model}, when initializing the pairing with a challenge-response protocol, this device who intends to share information with an AP sends a message~(\textit{Message 1}) for request. If the AP receives this message, it sends an acknowledgement message~(\textit{Message 2}) back. Finally, the IoT device receives \textit{Message 2} and feeds back the message~(\textit{Message 3}). However, during this process, an active attacker can launch two types of attacks to this AP and break the establishment:

\textbf{In a basic attack.} An attacker can impersonate the legitimate IoT device with the same MAC address and login credential. Then, it can be placed at different locations to send the same packets as the legitimate IoT device to the AP. If we lack a reliable secure scheme, this attacker would fool the AP to establish a malicious connection.

\textbf{In an advanced attack.} If an attacker is equipped with powerful hardware and process ability, the system would be easily attacked. In particular, a powerful attacker who has the priori knowledge of the protocol and the legitimate IoT device~(e.g., the MAC address, the coding scheme, the carrier frequency, the backscatter sequence, and signal strength) can successfully detect the initialization of the communication and estimate the channel state between itself and the AP. Then, based on the estimation and the priori knowledge, this attacker can craft a multi-path signal similar to the legitimate IoT device~\cite{fang2014you}. Finally, the attacker sends this well-designed signal with a directional antenna directly to spoof the AP. Accordingly, the AP receives the signal from the attacker and believe that this fake message is from a legitimate IoT device, which leads to unauthorized access to the AP.

\begin{figure}[t]
	\center
	\includegraphics[width=1.7in]{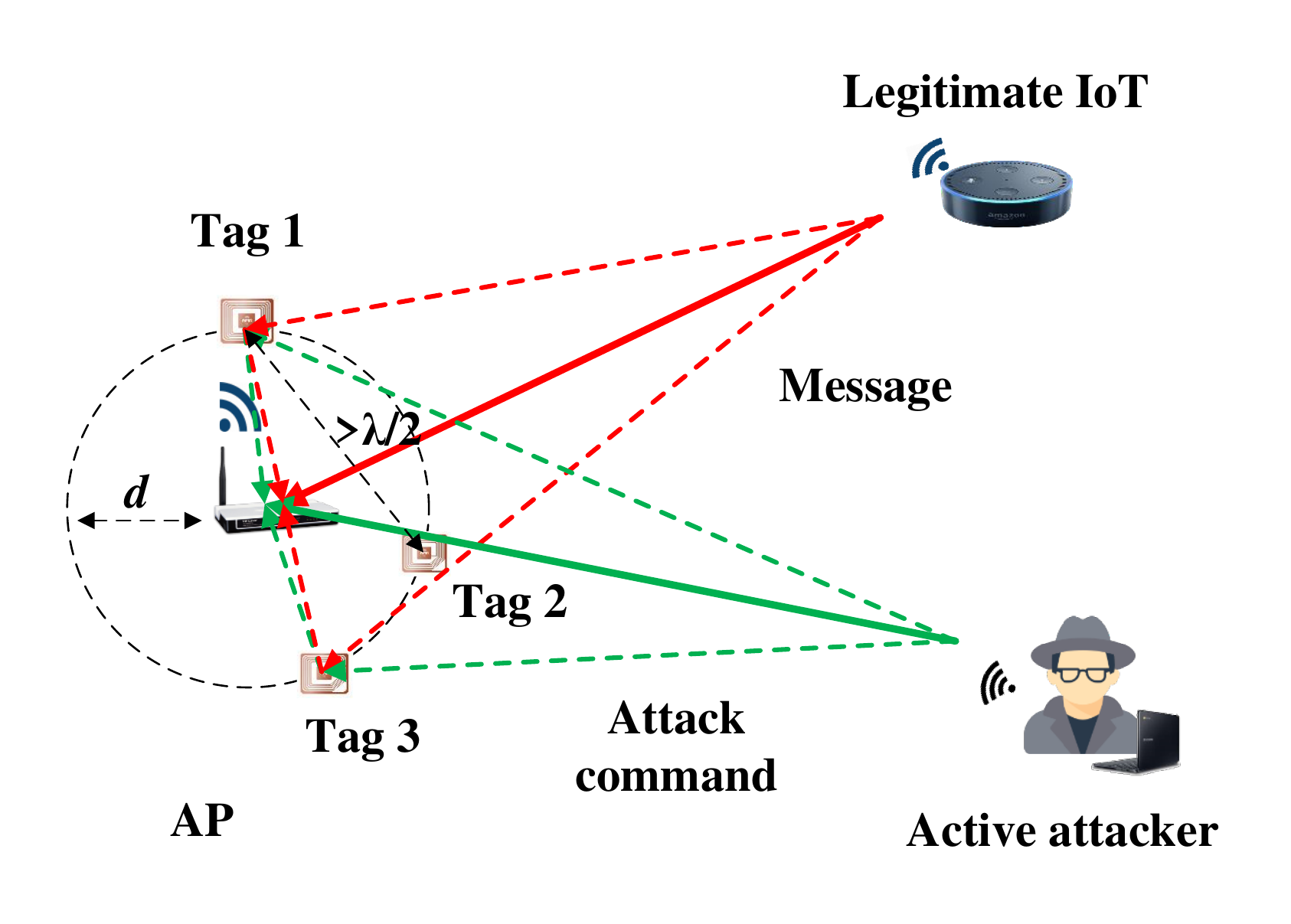} 
	\caption{We deploy three tags around the AP to defend against the active attacker. }
	\label{fig:place}\vspace{-0.3cm}
\end{figure}
ShieldScatter considers that the attacker will not be triggered to attack the IoT devices all the time. In other words, only if the attacker detects that a legitimate device is connecting or sharing data with the AP~(e.g., detecting \textit{Message 1}), the attacker will launch the attack. Besides, ShieldScatter only defends against active attacks and makes no exploration of protecting against passive attacks such as eavesdropping attacks and information leakage.

\subsection{ Propagation Signatures }\label{sec:signature}
\begin{figure}[t]
	\center
	\includegraphics[width=3.0in]{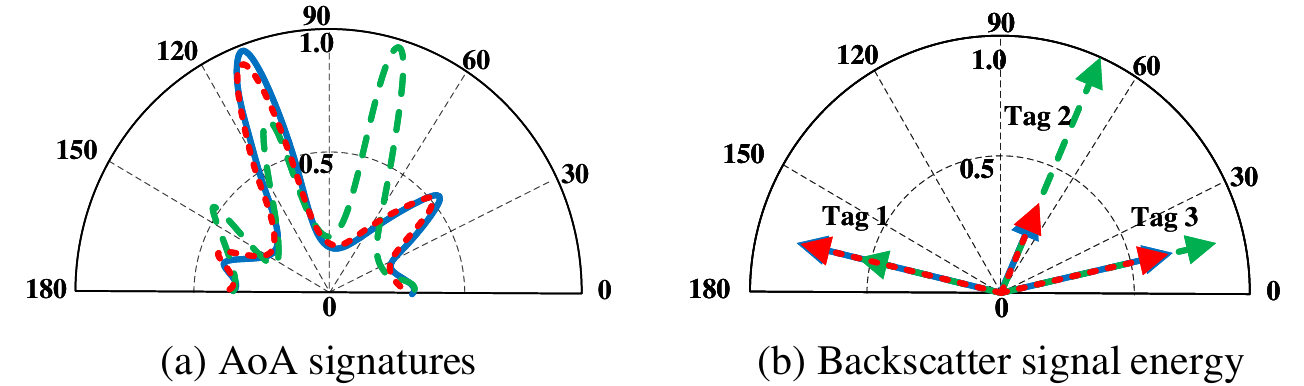} 
	\caption{If the signals are from the same devices, the AoA signatures have a strong similarity between them. Otherwise, the AoA signatures will be different. When we use three tags to create multipath, the average energy of each tags has the similar performance as AoA signatures. }
	\label{fig:AoA}\vspace{-0.3cm}
\end{figure}

Existing approaches to secure these active attacks by relying on extracting fine-grained propagation signatures from the physical-layer information. As shown in Fig.~\ref{fig:AoA}(a), SecureArray~\cite{xiong2013securearray} extracts the AoA signatures from the received signal with an antenna array. Then, based on the similarity of the AoA signatures, active attacks can be detected. However, devices may lack multiple antennas, and thus the methods of extracting fine-grained signatures with an antenna array will be inapplicable. Inspired by SecureArray that uses multi-path signatures to secure the devices, we observe that low-cost and battery-free backscatter tags can also generate such multi-path propagation signatures without using a large antenna array. According to~\cite{liu2013ambient}, backscatter communication is a new communication primitive where a tag transmits data by intermittently reflecting ambient signals. Specifically, when the transmitted signals resonate at the antenna of the tag, an AP can receive the signal consisting of the direct path and backscatter signal, which can be expressed as
\begin{equation}\label{eq:primer}
\\S_r(i)={S_t(i)+\alpha S_b(i)S_t(i)+\eta(i)},
\end{equation}
where $S_r(i)$ represents the received signal, $S_t(i)$ the transmitted signal from the legitimate IoT device, $\alpha$ the reflection attenuation coefficient, $S_b(i)$ the transmitted data bit of the tags, and $\eta(i)$ the ambient noise. We can easily find that the received signals are the combination of the source signal from the transmitter and the backscatter signal reflected by the tag.  Consequently, this backscatter signal  $S_b(i)$  can be treated as the multi-path signal, and thus this signal can serve as the distinguishing profile of the transmitter and defend against impersonate attacks.

In order to verify this idea, as shown in Fig.~\ref{fig:place}, we attach three tags around the AP, and the tags are separated at a distance larger than a half wavelength to mitigate the channel correlation. Besides, we place two USRPs in two separate positions to emulate a legitimate IoT device and an active attacker, respectively. Then, the legitimate IoT device is pairing with the AP based on a challenge-response scheme. At the same time, the AP controls the tags to reflect the signals in turn. According to~\cite{liu2013ambient}, since the backscatter tags are extremely sensitive to the signal strength and arrival direction, the backscatter signals will be significantly impacted if they are from different distances and directions. Consequently, we extract backscatter signals using the sliding windows and compare the average energy of each tag. The result is shown in Fig.~\ref{fig:AoA}(b).  We observe if these two pairing signals are from the same legitimate IoT device, the average energy of the tags will have a strong similarity (e.g., the red and blue dashed lines). Whereas, if one of the pairing signals is from an attacker, the distances, directions, and channel state between the tags and the IoT device differ from the attacker, and thus it will lead to significant differences with respect to the amplitudes of the tags (e.g., the green dashed line). This result presents a similar performance to AoA signatures and it motivates us to design ShieldScatter, a lightweight system to secure the IoT device by using backscatter tags. ShieldScatter intentionally creates multi-path signatures by controlling the tags to reflect the signals in turn. In order to construct a reliable profile to detect the threat, we also extracts other representative features and combines with a one-class SVM  classifier to identify the attackers.

Besides, we notice that the pairing or authenticating operation with a challenge-response scheme can be completed within the coherence time. Though the wireless channel is easily affected by the environmental states, however, within coherence time, the channel can be treated as stable even in dynamic environments. The coherence time can be defined as $T={{9\lambda}\over{16\times\pi\times{{v}}}}$, where $ \lambda$ represents the carrier wavelength and $v'$ indicates the maximum velocity of a legitimate IoT devices~\cite{xiong2013securearray,steele1999mobile}. Therefore, when the people or the devices are moving at a slow speed, we can still guarantee the similarity of the backscatter signals.
\begin{figure}[t]
	\center
	\includegraphics[width=2.3in]{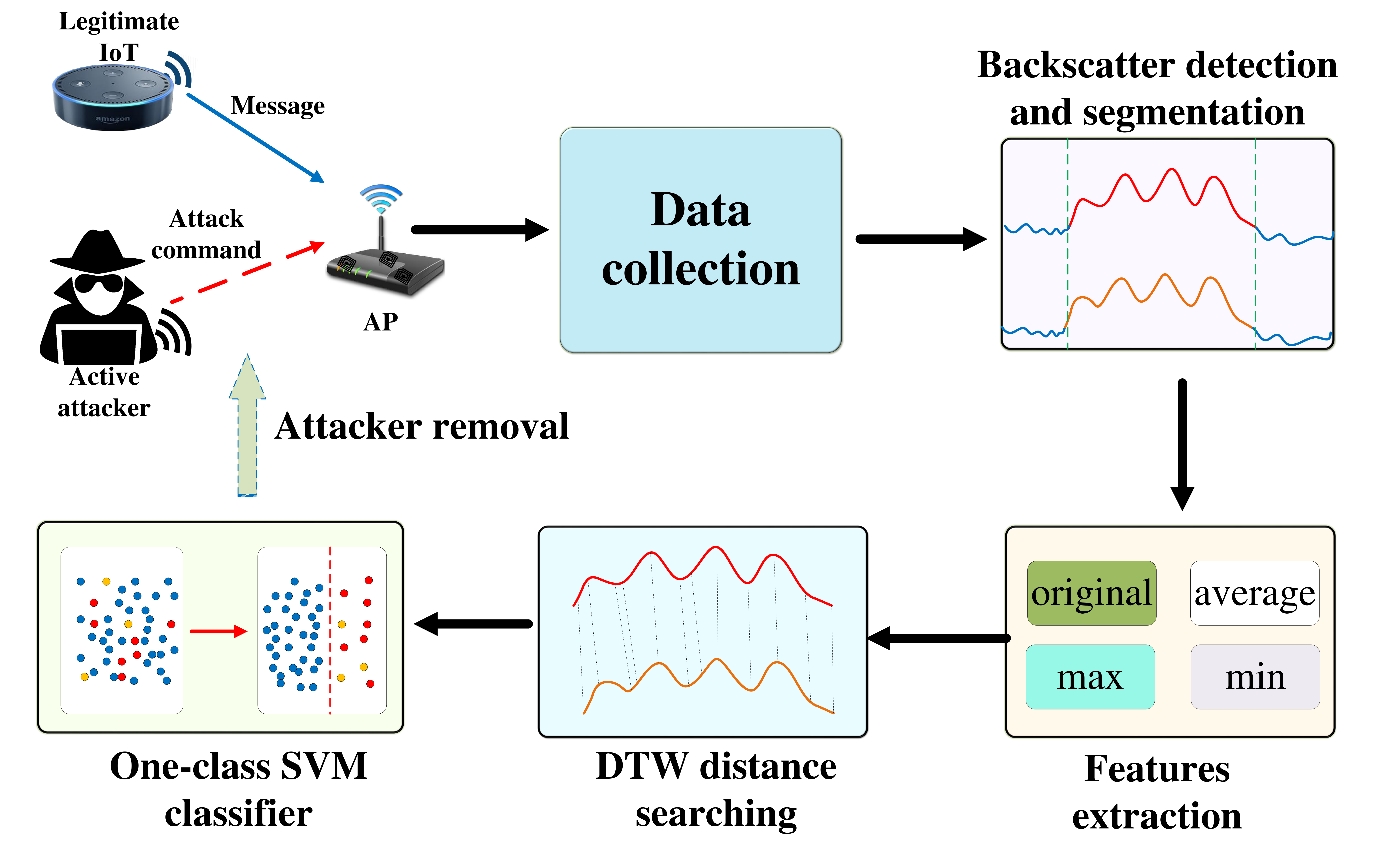} 
	\caption{System overview.}
	\label{fig:sysetm}\vspace{-0.3cm}
\end{figure}

\section{System Design}\label{sec:system}
In this section, we first present an overview of ShieldScatter which consists of four key steps. Then, based on a basic attack, we elaborate on each step and provide the technical details in the following subsections. Finally, we provide a detailed description of advanced attack and design a tag-random  algorithm and combine multiple receivers to defend again this attack.
\subsection{ System Overview }\label{sec:overview}
The basic idea of ShieldScatter is to construct sensitive multi-path propagation signatures using several backscatter tags attached around the AP instead of a large antenna array. In particular, as shown in Fig.~\ref{fig:attack_model}, when the legitimate IoT device is pairing with the AP, it can easily be attacked by a fake transmission. To defend against this attack, upon detecting the suspicious transmission, the AP is asked to control the tags to work in turn during this process. Then, by comparing the signatures from the same device (e,g.,  \textit{Message 1} and \textit{Message 3}) or from different devices (e,g., \textit{Message 3} and the suspicious command), ShieldScatter triggers the security system to verify whether the suspicious messages are from the legitimate device and then defend against active attacks.

To detect the attacks, as illustrated in Fig.~\ref{fig:sysetm}, at a high level ShieldScatter needs to go through the following four steps. First, based on the collected data at the AP, ShieldScatter detects and segments the signals that include the backscatter signal. Second, ShieldScatter extracts representative features from the segments. Third, to construct a reliable propagation profile, ShieldScatter compares the features by computing the distances with DTW. Finally, based on the profiles with respect to the DTW distances, ShieldScatter can identify and defend against the attacks with a one-class SVM classifier. In the following subsection, we elaborate each step based on a basic attack where an active attacker can be placed various locations for attacking. As for the advanced attack, it will be discussed in Section~\ref{sec:powerful attacking}.

\subsection{ Backscatter Detection and Segmentation }\label{sec:segmentation}


\begin{figure}[t]
	\center
	\includegraphics[width=2.3in]{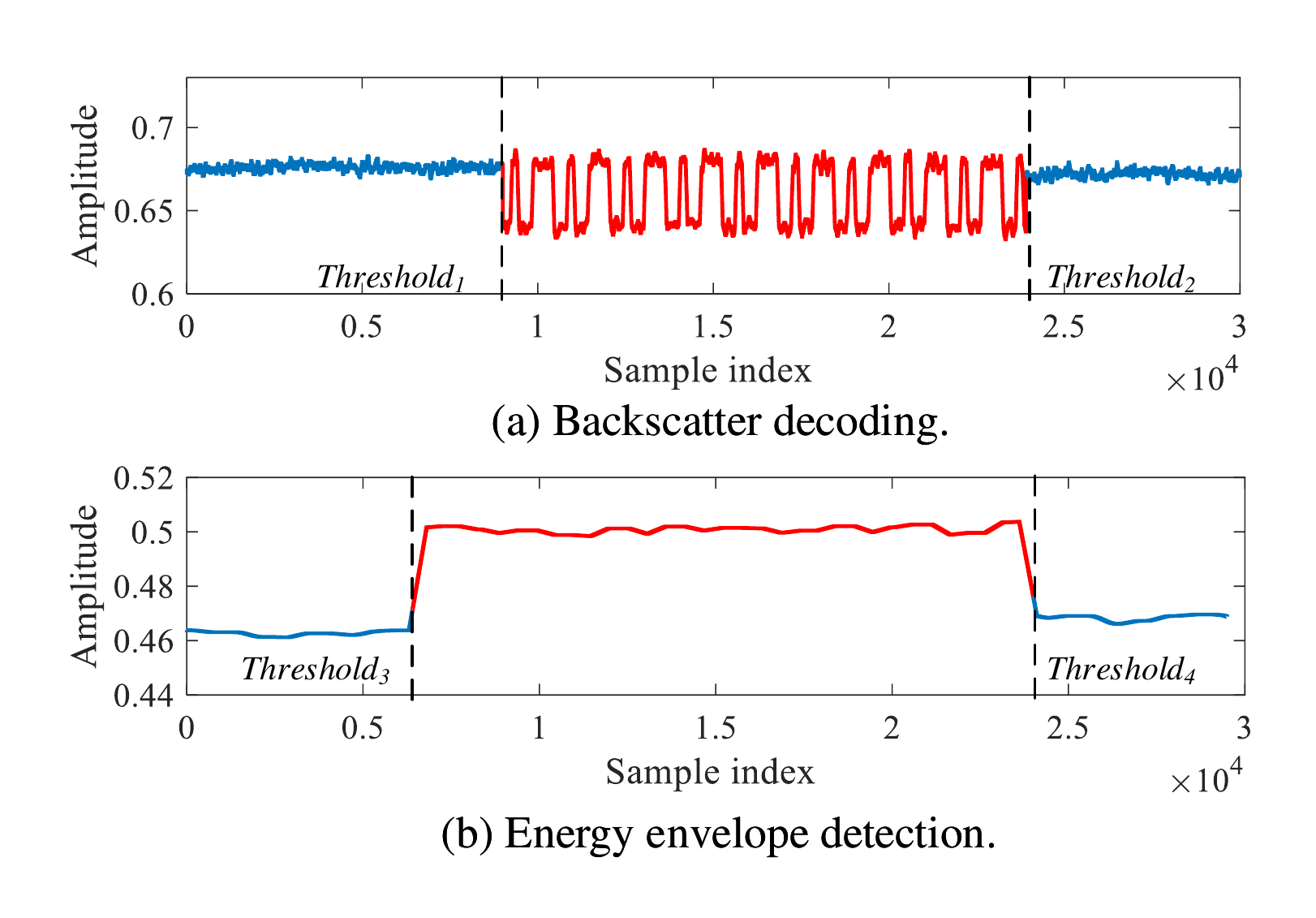} 
	\caption{ShieldScatter detects and segments backscatter component by combining backscatter decoding and energy envelope detection.}
	\label{fig:segment}\vspace{-0.3cm}
\end{figure}

\begin{figure*}[t]
	\center 
	\subfigure[Original signals.]
	{ \includegraphics[width=4.50cm]{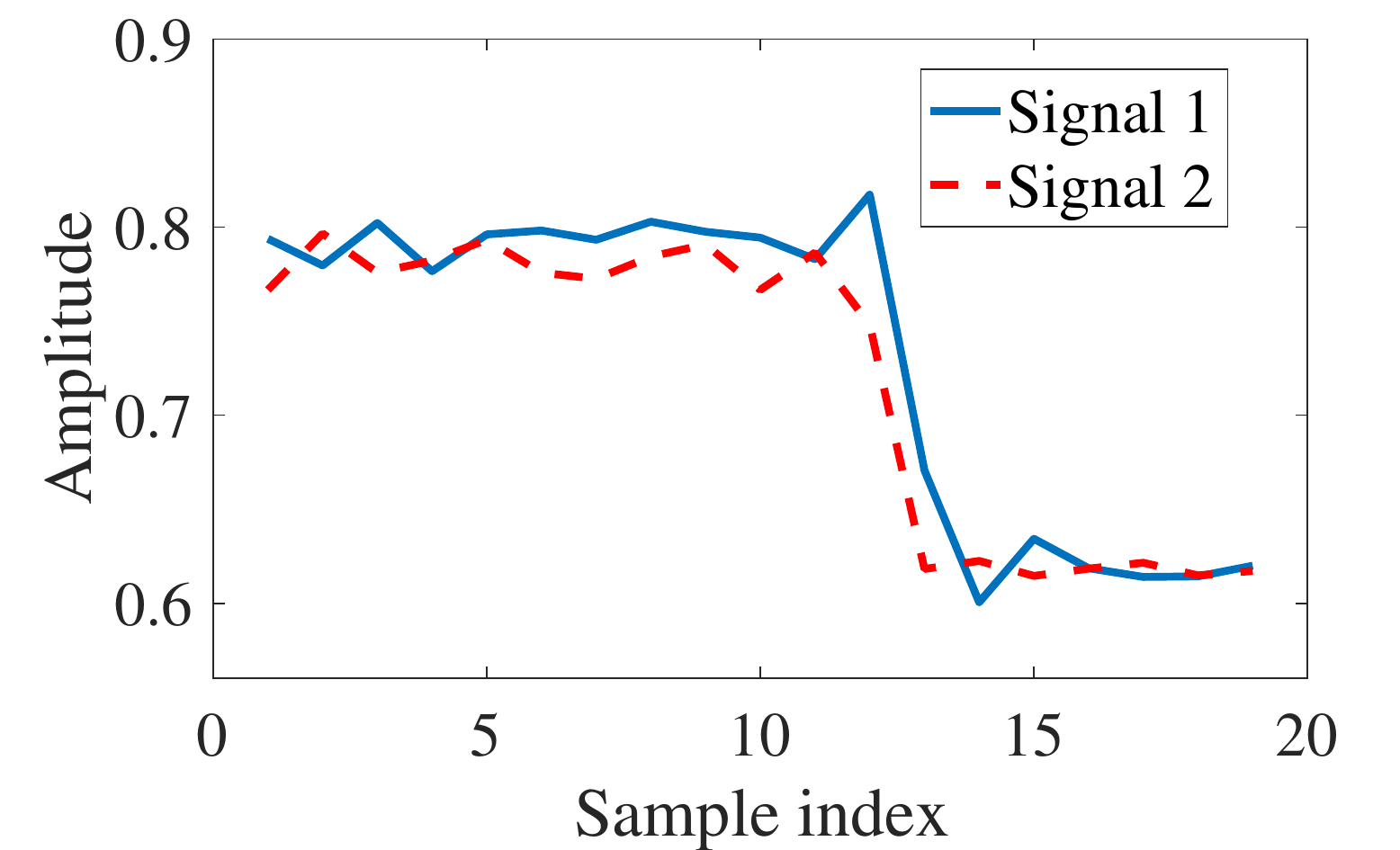}  }
	\subfigure[Smoothed signals.]
	{ \includegraphics[width=4.50cm]{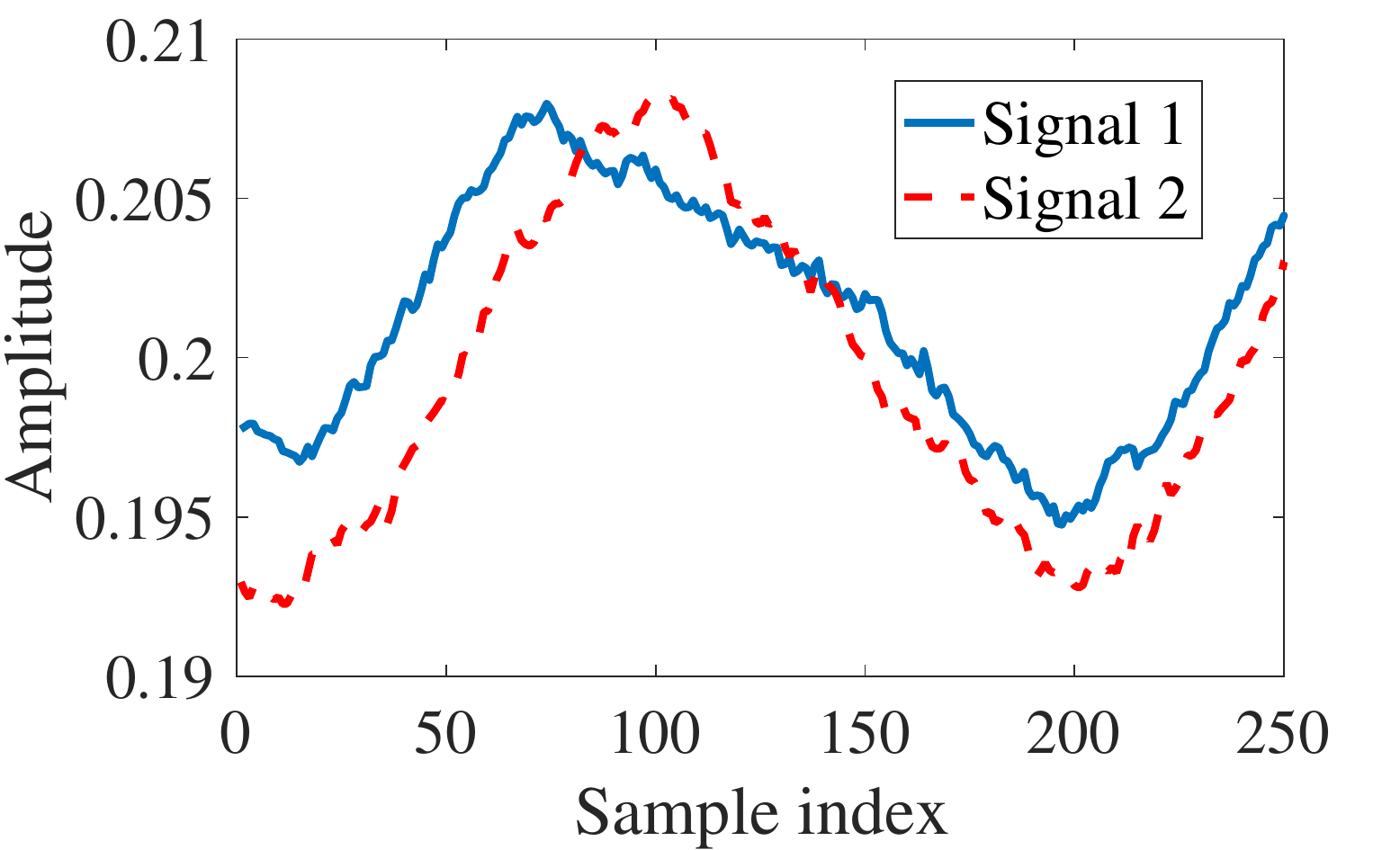}  }
	\subfigure[Maximum.]
	{ \includegraphics[width=4.5cm]{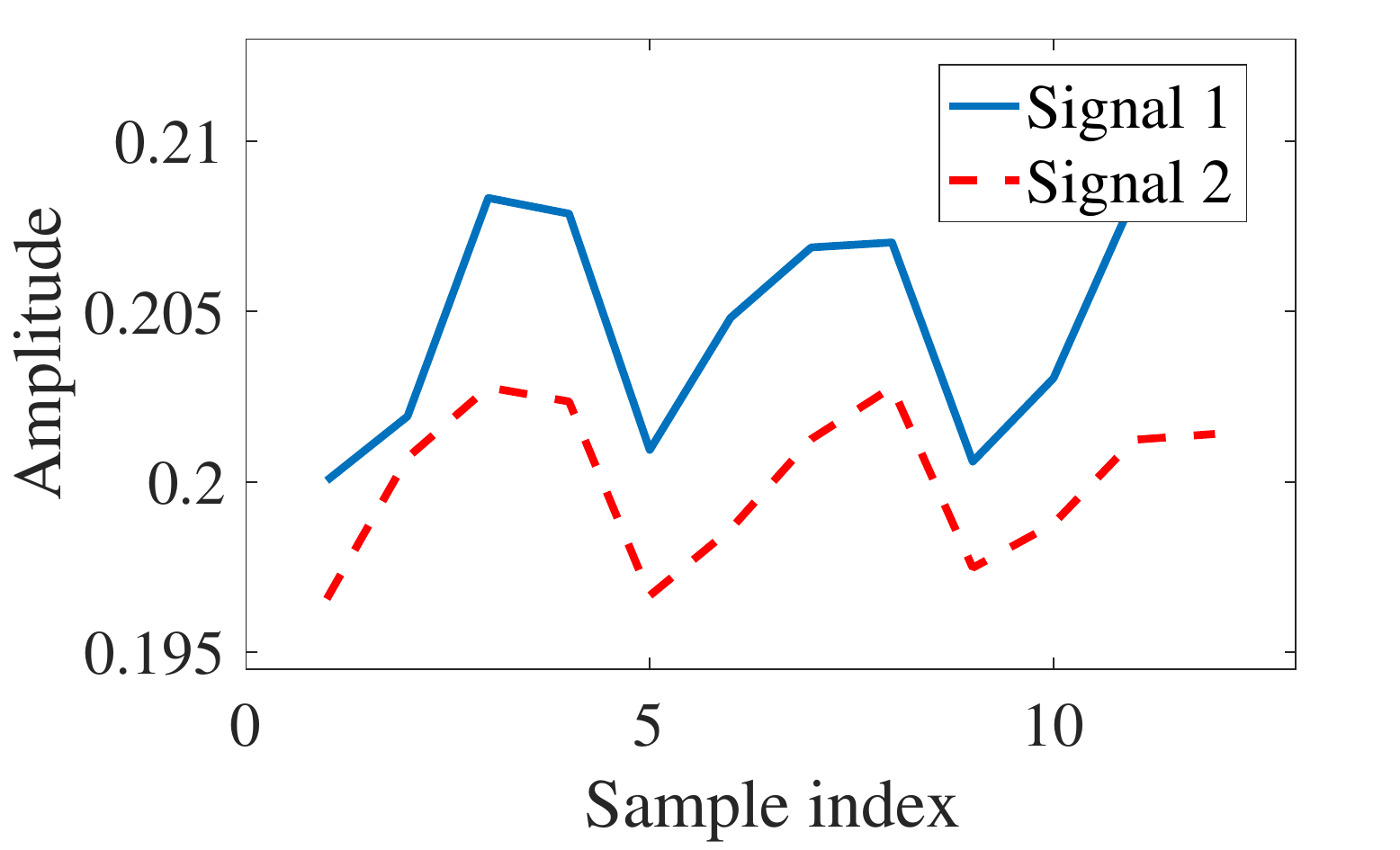}  }
	\caption{Features extracted from the same device.}
	\label{fig:feature_positive}\vspace{-0.3cm}
\end{figure*}

\begin{figure*}[t]
	\center 
	\subfigure[Original signals.]
	{ \includegraphics[width=4.5cm]{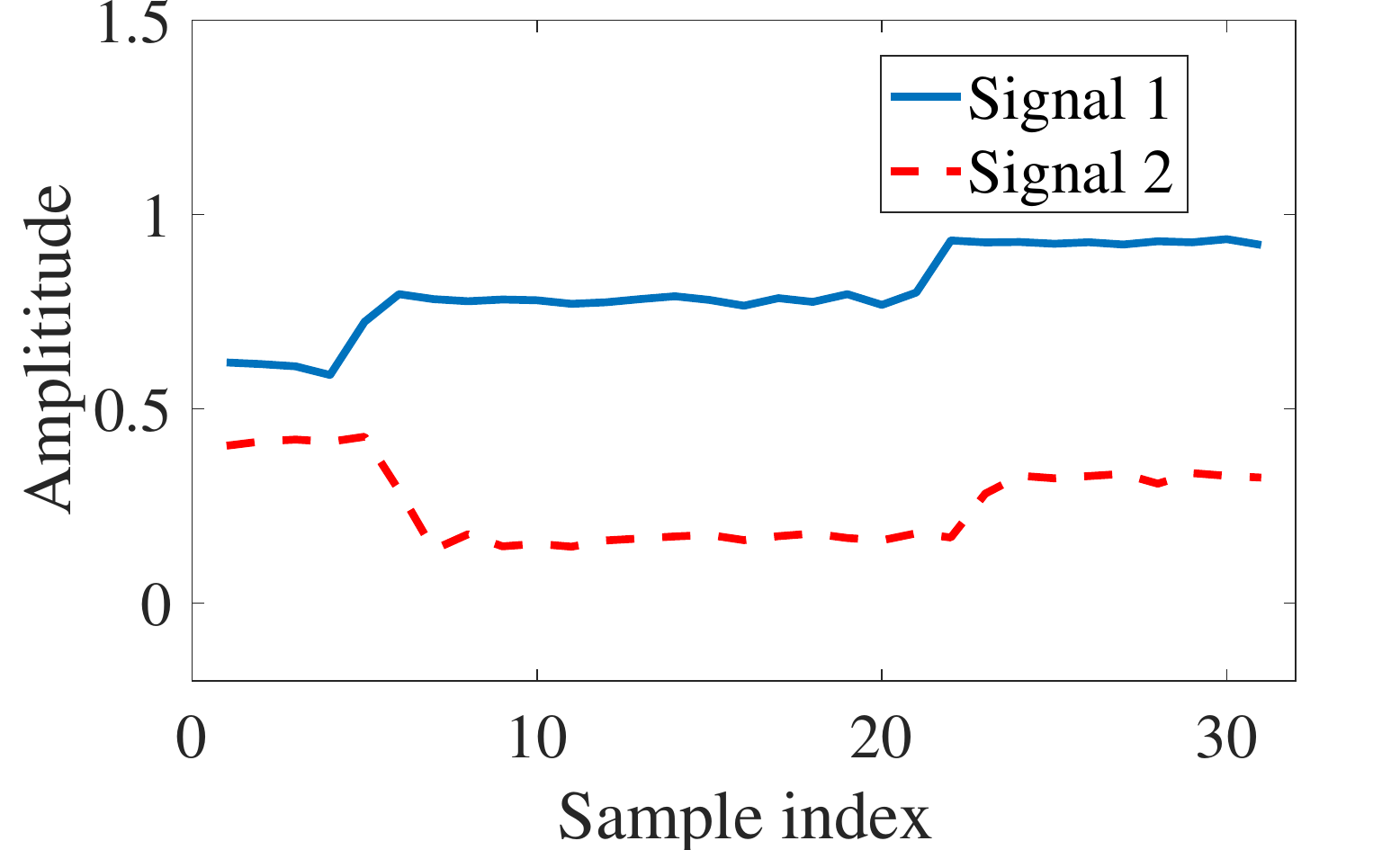}  }
	\subfigure[Smoothed signals.]
	{ \includegraphics[width=4.5cm]{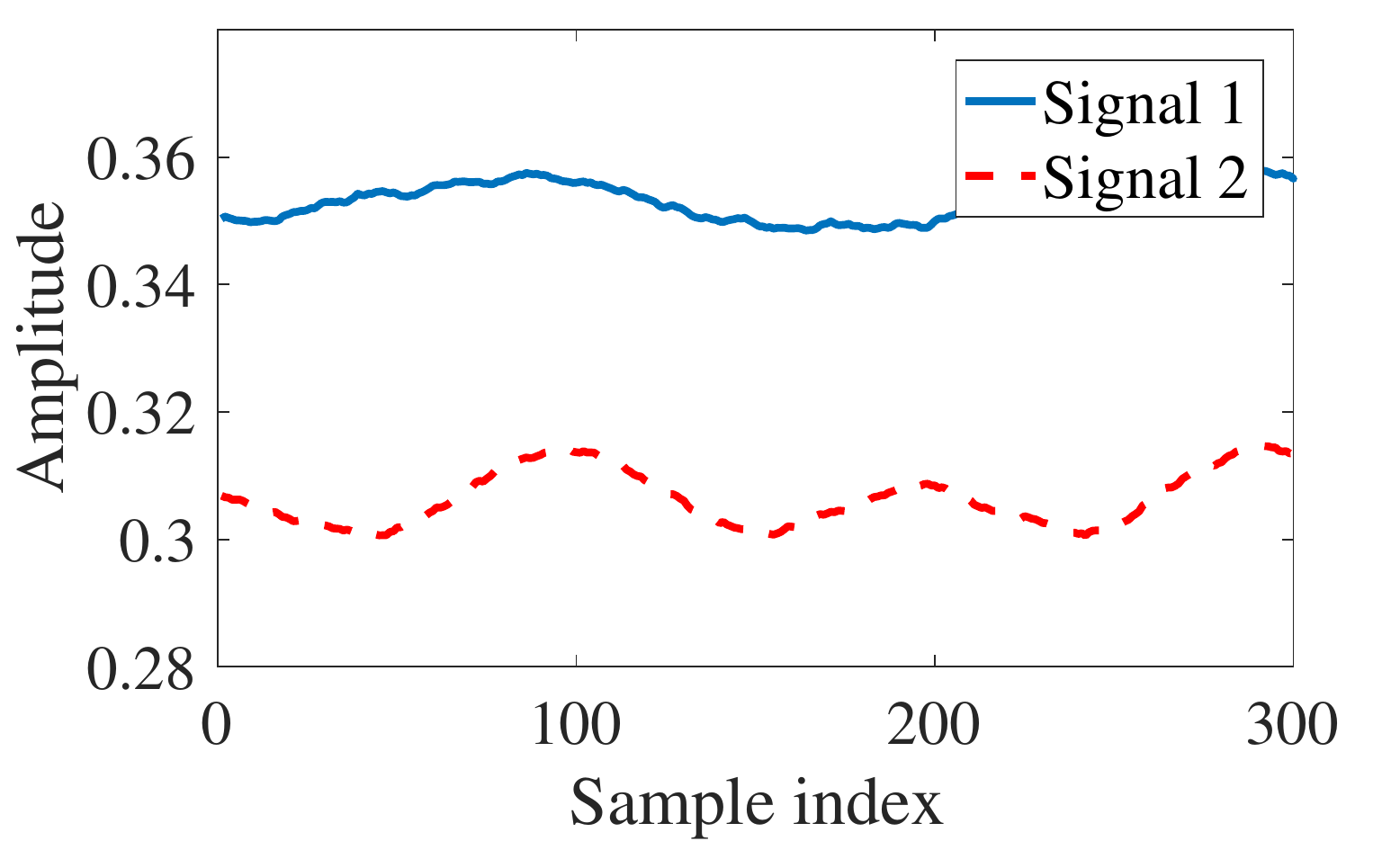}  }
	\subfigure[Maximum.]
	{ \includegraphics[width=4.5cm]{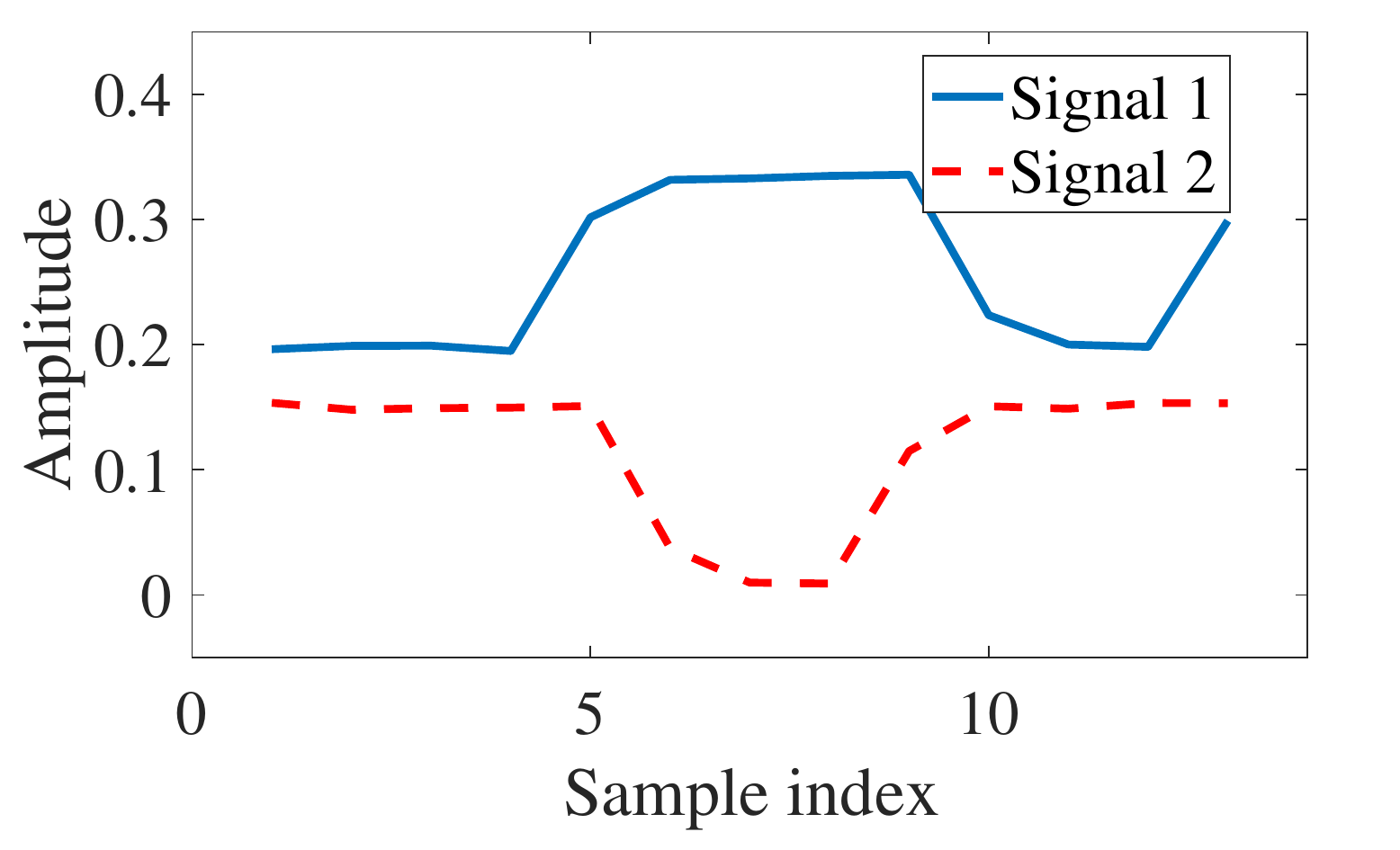}  }
	\caption{Features extracted from the different devices.}
	\label{fig:feature_negative}\vspace{-0.3cm}
\end{figure*}
Recall that ShieldScatter constructs sensitive multi-path propagation signatures by controlling the backscatter tags to work in turn. ShieldScatter needs to detect and segment the received signals that contain the backscatter component for the first step. In particular, we first decode the received signals using a moving average method as in~\cite{liu2013ambient} where we use a sliding window with a length of 50 samples to smooth the signal. After smoothing, ShieldScatter decodes the message of backscatter signals as shown in Fig.~\ref{fig:segment}(a). In order to determine the segment, ShieldScatter follows the principle: if the AP can continuously detect the backscatter signals, then the corresponding original signals samples from the starting point to the ending point are considered that contains the backscatter signal. We mark the starting point and the ending point as $\eta_1$ and $\eta_2$, respectively.

After that, ShieldScatter can achieve a raw signal segmentation to detect the backscattered signals. However, because of the imperfect circuit design and noise, it is not accurate enough to segment the signals. Thus, to improve the accuracy of the signal segmentation, we employ an energy envelope detection algorithm for assistance. Specifically, a sliding window upon the received signal amplitude is used to detect the backscatter, where we calculate the average energy $E(i)$ within this sliding window by
\begin{equation}
\\E(i)={{1\over N}{ \sum_{i=1}^{i+N}{|x(i)|^2}}},
\end{equation}
where $N$ is the length of the sliding window, and $x(i)$ is the amplitude of the sample at index $i$. After calculating the energy of the signals, we can easily yield the energy envelope as shown in Fig.~\ref{fig:segment}(b). It is obvious that the energy envelope changes greatly when the backscatter tags are working. Besides, we also find that the backscatter signal is alway located at the center of samples without backscatter, which inspires us that the energy envelope will experience a large variance when backscatter is working. Thus, in order to determine the starting and ending points of the segment that contains the backscatter signal, we calculate the variance of the energy envelope by
\begin{equation}
\\V(j)=Var[E(j):E(j+N) ],
\end{equation}
where $Var[E(j):E(j+N) ]$ represents we calculate the variance in every $N$ samples and $ V(j)$ represents the variance at index $j$. Then,  we mark the starting point as $\eta_3$ and ending point as $\eta_4$ to determine the backscatter signal with the following two constraints; when $0<j<\eta_3$, we need
\begin{eqnarray}
& & V(j)<T,V(\eta_3+1)>T, 
\end{eqnarray}
and for any $\eta_4<j<M$,
\begin{eqnarray}
& & V(j)<T,V(\eta_4-1)>T,
\label{eq:diaggonaldominance}
\end{eqnarray}
where $M$ represent the total number of $V(j)$ and $T$ is the dynamic threshold. According to our experimental study, we set the threshold as $e^2$ where $e$ is the minimum energy of all the tags and it can be obtained by the backscatter decoding. Based on the threshold, we can obtain the starting and ending points as $\eta_3$ and $\eta_4$, respectively.

Finally, we combine the results of backscatter decoding and energy envelope detection to determine the segment by
\begin{eqnarray}
\eta_s={\left.(\eta_1+\eta_3)\right/2},
\eta_e={\left.(\eta_2+\eta_4)\right/2} ,
\end{eqnarray}
where $\eta_s$ and $\eta_e$ represent the final decision for the starting and the ending points of the segment, respectively. 

\subsection{Feature Extraction}\label{sec:features}
Before constructing reliable multi-path propagation profiles, ShieldScatter should extract representative features from the segments. According to the ambient backscatter theories~\cite{liu2013ambient}, the backscatter signal is an additional multipath generated by the tag. This additional multipath can either constructively or destructively interfere with the ambient signal. All of these multipaths mainly affect the amplitude of the ambient signal. Thus, to construct reliable signatures from the obtained segments, ShieldScatter selects the features with respect to the signal amplitude.

Intuitively, since the location, the tag reflection, and the channel sate of the IoT will affect the propagation signatures, ShieldScatter can directly use the received raw data (we call it original signals in this paper) for comparison. However, the existing ambient noise will lower the similarity of the comparison. Besides, the strong direct path signal would also overwhelm the fine-grained backscatter signatures. Therefore, only the original signal is not enough to construct a reliable propagation profile. Consequently, in our system, besides \textit{original signal}, five other significant features with respect to the signal amplitude, including \textit{smoothed signal}, \textit{energy envelope}, \textit{variance}, \textit{maximum} and \textit{minimum}, are also extracted to construct our unique and sensitive multi-path propagation profile for the legitimate IoT device. Specifically, because radio propagation from various positions and tags will lead to different power at the receiver, we extract the energy envelop of the signal as the feature. Besides, since the frequency of the backscatter signal is much lower than the signal directly from the IoT, we extract it with a smoothed mechanism, and thus smoothed signal represents the feature of the backscatter signal. In order to acquire smoothed signals, ShieldScatter filters the original signals by using a sliding window. Furthermore, as variance, maximum and minimum can also reflect the backscatter variations caused by the tag hardware and radio propagation, ShieldScatter extracts these features by computing the variance, maximum, and minimum using every 50 data samples of the original signals. Accordingly, ShieldScatter can obtain six \textit{feature series} for each of the segments obtained in Section~\ref{sec:segmentation}. As shown in Fig.~\ref{fig:feature_positive} and Fig.~\ref{fig:feature_negative}, the feature series are extracted from the signals of the same and different devices within coherence time, respectively. It is obvious that the features are similar when the signals are from the same devices. Otherwise, if the signals are from different devices, the features extracted from these two signals are quite different. Accordingly, these representative features can be used to construct a reliable profile and secure the legitimate IoT device.

\subsection{DTW Distance Searching}\label{sec:DTW}
After acquiring the feature series from the segments, ShieldScatter needs to compare the features to detect active attackers. Intuitively, the simplest approach is to calculate the correlation of every two corresponding feature series directly. However, because of the noise and imperfect circuit design of the backscatter tags, even though the schemes proposed in Section~\ref{sec:segmentation} have been exploited to detect and segment the signal, they still cannot guarantee an absolutely accurate partition of the received signal. Besides, since the tags reflect the signal in a periodic way, the imperfect segmentation of the signals leads to a shifting of the features, which can be seen in Fig.~\ref{fig:feature_positive}. Thus, simply computing the correlation to compare the feature series will lower the similarity and are not applicable to our system. Instead, a approach that can mitigate the unfavorable effects caused by misalignment is more desirable for comparison in our design.


Inspired by PinIt~\cite{wang2013dude} and the method for the word matching in speech recognition that have a similar nature of signature shifting, we find that the most commonly used method DTW can be adopted to mitigate the effect of misalignment. Thus, we compare the similarity of the features to construct the propagation profiles by using DTW distance computing~\cite{salvador2007toward} as follows:
supposed given the two feature series $X(i)$ and $Y(j)$ of the corresponding segments (e.g., the feature \textit{maximum} of \textit{Message 3} and the suspicious message in Fig.~\ref{fig:attack_model}), the goal of DTW is to find the minimum cost of the mapping sum from the feature series $X(i)$ to $Y(j)$, which is defined by using the Euclidean distance
\begin{equation}
\\w(i,j)=|X(i)-Y(j)|.
\end{equation}
Then, based on the Euclidean distance between each two samples of these two feature series, a dynamic programming algorithm is used for DTW to search for the warp path distance. 
Accordingly, We define DTW in mathematical expressions as 
\begin{eqnarray}
& \underset{W}{\text{min}} & \sum_{j=1}^{m} \sum_{i=1}^{n}   w(i,j) \\
& \text{s.t.} & sp=w(1,1),{ep}=w(m,n), \notag \\
& & {st}(i)\le st({i+1}),st(j)\le{st}({j+1}).
\end{eqnarray}
where $W$ represents the route matrix, $sp$ and $ep$ the starting-point and ending-point, respectively. $st(i)$ indicates the horizontal axis coordinates at the $i_{th}$ step, and the two constraint conditions have guaranteed the boundary and monotonicity for the route selection in DTW.

In our system, in order to reduce computational complexity, we compute the DTW distances by dividing the feature series into different chunks instead of directly using the entire original signal and smoothing signal. ShieldScatter segments the feature series equally into 128 chunks. Then, ShieldScatter computes the DTW distance in each corresponding chunk. As for energy envelope, variance, maximum and minimum, ShieldScatter divides them into 58 chunks and computes the DTW distances, respectively. Accordingly, we can finally obtain a propagation profile by combining all the DTW distance results, which is a vector with the size of 488. Compared with the method of calculating correlation directly, DTW mitigates the effects caused by misalignment.

\subsection{One-Class SVM Classification}\label{sec:SVM}
Based on the similarity comparison of the extracted features, we obtain a propagation profile vector with the size of 488 for every DTW processing. As mentioned, the signals from same device will experience the same multipath caused by the intentionally deployed tags, which will lead to high similarity and short DTW distance for the features. Otherwise, the DTW distances will be significantly different. Thus, we can transform our problem of detecting the suspicious signals into the problem of distinguishing the propagation profile vectors so as to defend against the spoofing signals.

At an intuitional level, in order to distinguish the propagation profile vectors, the most likely method is to set fixed thresholds for each value in the vector. Then, if all the values are lower than the corresponding thresholds, the signals can be considered as positive samples. However, this method is unreliable, since the received signals will be significantly affected by the environment noise  in dynamic environments, which make it difficult to determine these fixed thresholds.

\begin{figure}[t]
	\center
	\includegraphics[width=2.0in]{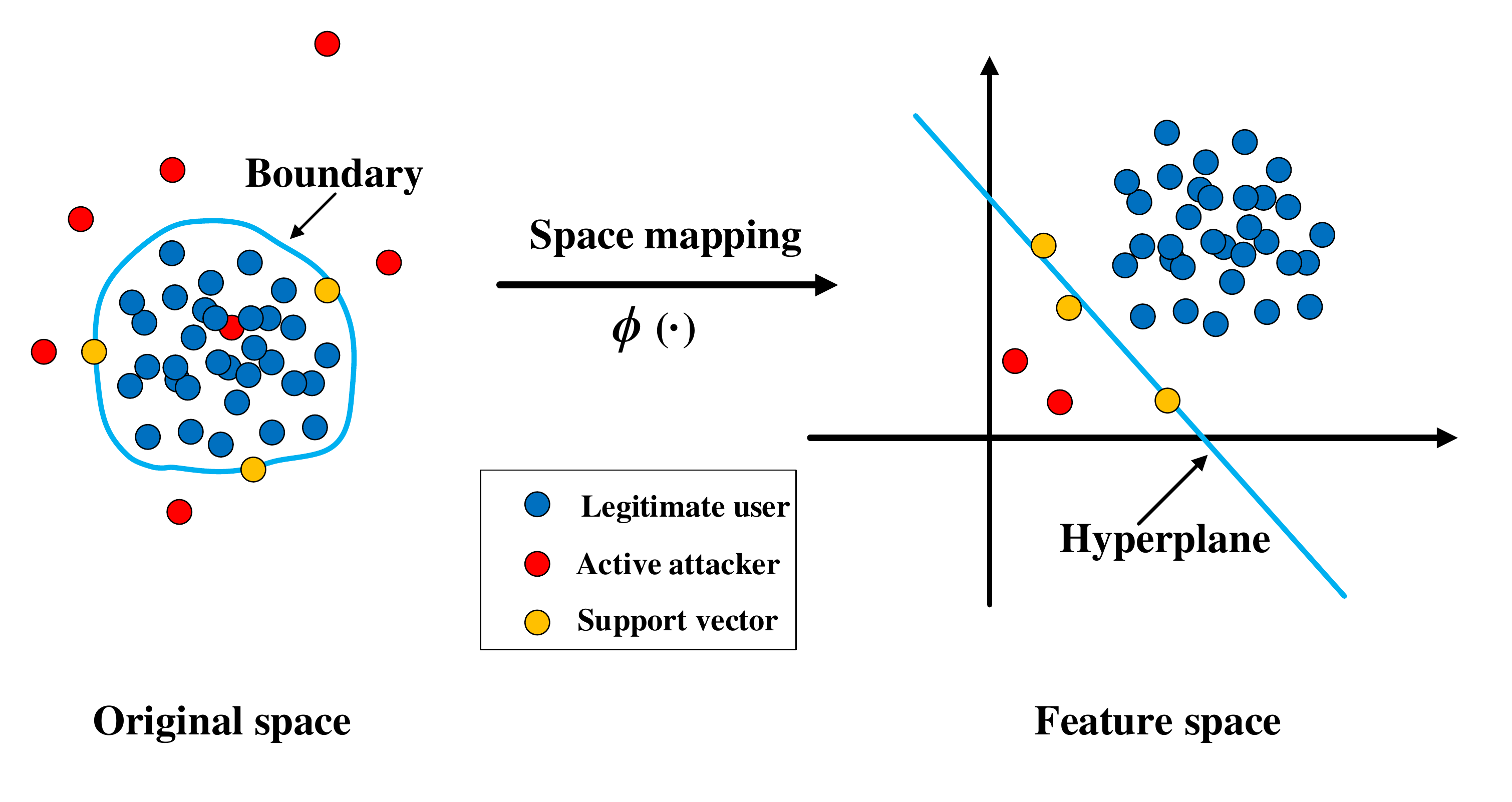} 
	\caption{The profiles of the legitimate IoT device can be mapped into a feature space and separated from the attacking samples.}
	\label{fig:SVM}
\end{figure}

To distinguish the legitimate IoT device and attacker profiles, ShieldScatter formulates our problem as a classification model. In particular, given a large number of training profiles as $[\bf{x}_1, \bf{x}_2, ...\bf{x}_\emph{i}... \bf{x}_\emph{l}] $,
where $l\in$ $\mathbb{N} $ is the number of profiles, and $\bf{x}_\emph{i}$ the profile vector of the profile $i$. The size of each propagation profile vector with respect to the DTW distances is 488. As shown in Fig.~\ref{fig:SVM}, if the profiles are collected from the positive samples, the profiles of the legitimate IoT device will be similar and they will gather together closely. Reversely, the samples that are from the attackers will be further away from these positive profiles except for a few outliers that have short DTW distances. Thus, the goal of ShieldScatter is to find an optimal boundary to capture most of the positive samples and exclude the negative samples. 

In order to define the optimal boundary, we formulate our system as a one-class support vector machine (SVM) model, where we can map the sample from the original space into a feature space using function $\phi(\bf{x})$~\cite{scholkopf2001estimating}. In the feature space, the samples can be separated by a hyperplane with the maximum margin method, where this hyperplane is defined by some samples called support vectors in the training set. In order to seek out these support vectors, this problem can be formulated as 
\begin{eqnarray}
& \underset{\alpha}{\text{min}} & {{1\over 2 } {\sum_{ij}   {\alpha}_i{\alpha}_j \bf{\emph{k}}(\bf{x}_\emph{i},\bf{x}_\emph{j})  }}, \\
& \text{s.t.} & {0\le{\alpha}_i\le{1\over{vl}   }},  { \sum_{i}{\alpha}_i=1} 
\end{eqnarray}
where $v\in$  (0,1] is an upper bound on the fraction of the outliers and a lower bound on the
fraction of support vectors. $k(.)$ represents the Gaussian kernel, which is defined as 
\begin{equation}
\\\bf{\emph{k}}(\bf{x}_\emph{i},\bf{x}_\emph{j}) ={{\phi}(\bf{x}_\emph{i})}\times{{\phi}(\bf{x}_\emph{j})  } ,  
\end{equation}
Then, the decision function of ShieldScatter is defined as
\begin{equation}
\\f(x)=\text{sgn}({\sum_{i}   {\alpha}_i\bf{\emph{k}}(\bf{x}_\emph{i},\bf{x})  }-\rho),
\end{equation}
where $\bf{x}$ is the sample of the testing profiles, $\bf{x}_\emph{i}$ the $i_{th}$ support vector and $ \rho$ is the function bias. Hence, based on the hyperplane decided by the obtained support vectors, the testing profiles can be classified into either legitimate IoTs or attackers.

\subsection{Advanced Attack Defense}\label{sec:powerful attacking}
In an advanced attack, an AP would be attacked by a powerful attacker who has the strong priori knowledge of the system, including the carrier frequency, the  modulation and coding schemes, the bitrate of the backscatter tags, and so on. With the knowledge, she can easily impersonate the legitimate IoT device to spoof the AP. Specifically, we consider that a legitimate IoT device is pairing with an AP. At the same time, we control the tags to reflect the signals and protect this system, and thus this reflection can be treated as a new path from the tag. According to~\cite{fang2014you}, the signals propagate in the air through multiple paths due to indoor obstacle reflection, diffraction, scattering and tag reflection. Therefore, each multi-path signal can be considered as a delayed signal copy
	\begin{equation}\label{eq:pow}
	\\{\bf{Y}}_n={ \sum_{n}  {\beta}_n {\bf{X_\emph{n}}} ({t}- {t}_\emph{n})  },
	\end{equation}
where ${\bf{Y}}_n$ is the received signal from the transmitter, ${\bf{X}}_n$ the delayed signal copy caused by the $n_{th}$ path, $ {\beta}_n$ the attenuation factor, and ${t}_\emph{n}$ the time delay of the path $n$. Thus, the received signal is the superposition of the indoor and the tags' reflection.  If the order of the tags' reflection is fixed, the multipath caused by the tags and indoor environment is predictable. Then, based on a training sequence~\cite{fang2014you}, the attacker can estimate the attenuation factor, the time delay of each path, and also the real channel impulse response from the attacker to the IoT device. Accordingly, an attacker could well design its transmitted data,  craft a virtual multi-path signal, and directly send it to the receiver with a well-estimated power. Consequently, the receiver would receive the signals that are similar to those from the legitimate IoT device, and believe that the message is from this device. Thus, even though the attacker is placed in any location, it is still able to spoof the receiver with this  virtual multi-path signal.

\textbf{Tag-random scheme.} To defend against such a powerful attacker, we design a tag-random scheme to improve security. In particular, ShieldScatter first controls the tags to reflect source signals in a fixed order when receiving the pairing message (e.g. \textit{Message 1}). After that, if receiving a suspicious message (e.g. \textit{Message 3}), ShieldScatter reflects this message with a random order and records this order. Next, we decode the backscatter signals from the tags and divide them into multiple segments by the different backscatter tags. Finally, the receiver can rearrange these segments with the same order in \textit{Message 1} and operate \textit{feature extraction, DTW distance searching, and one-class SVM classification}. Based on this scheme, an attacker may successfully estimate the actual multi-path channel characteristic, and create a well-designed multi-path signal following the backscatter order in \textit{Message 1}. However, each time, the AP controls the tags to reflect the signal (\textit{Message 3}) with a random order, and thus the attacker cannot predict the right order of the tags. Therefore, the attacker is unable to estimate and emulate the real channel impulse response and fails to attack this pairing process. Therefore, the attacker is unable to estimate and emulate the real channel impulse response and fails to attack this pairing process. In order to improve the security, we can also use a voting authentication scheme, where we slightly modify the authentication protocol and require the IoT device to authenticate several times with a voting scheme. In this case, we can also protect the legitimate IoT device when it is detected as an attacker.

\textbf{Multiple receivers combination scheme.} In a home scenario, as shown in Fig.~\ref{fig:sketch}, there will be multiple devices at the same time. When an IoT device is pairing or sharing data with AP (R), the signals will be reflected by the tags around this receiver. Accordingly, besides AP (R), all the other APs in the room can also receive the message and backscatter signals. Thus, to improve the accuracy, all the APs can construct similar profiles from the received signals and combine them to identify the attacker. In particular, when a device is transmitting data to AP (R), we control the tags to reflect the signals in turn. At the same time, in addition to AP (R), all the other APs in the room also record this signal (e.g., \textit{Message 1}). After that, if hearing the suspicious message (e.g., \textit{Message 3}), similar to AP (R), all the APs also collect this signal from the IoT device. Since the signal is reflected by the tags around AP (R), the backscatter signal can also be received by all these receives. Then, as mentioned before, all the APs will segment the received signals and extract the features from the backscatter signals respectively. Accordingly, all the APs can match the features and identify whether the suspicious message is from an attacker with the SVM classifier. Finally, we will combine all the results to improve the detecting rate with a voting scheme. Based on this scheme,  even though the legitimate IoT device is attacked by a powerful attacker, it is difficult for this attacker to estimate and design a crafted virtual multi-path signal when increasing the number of APs. Thus, we can successfully defend against an active attacker.
%

\begin{figure*}[t]
	\center
	\includegraphics[width=5.5in]{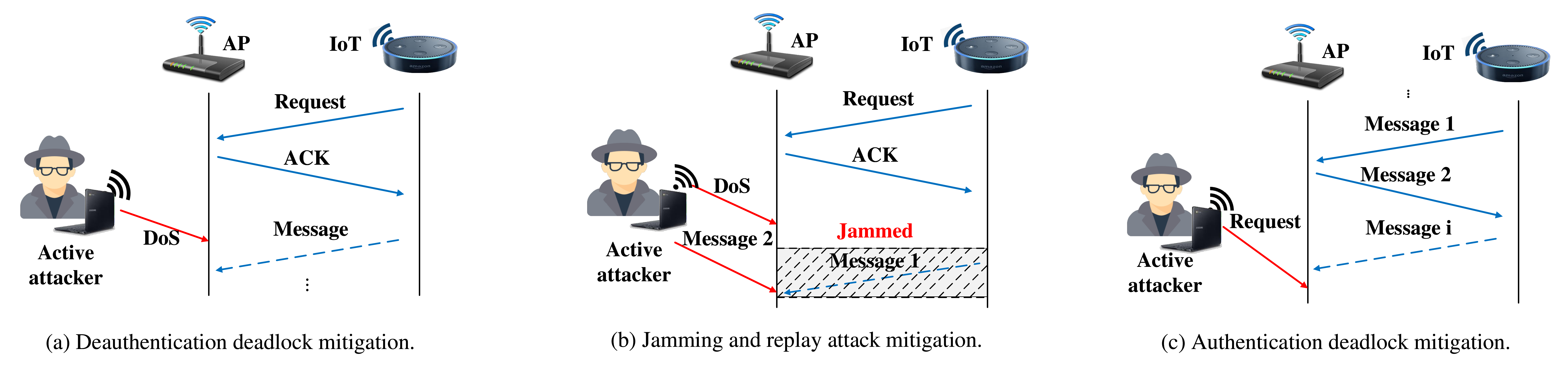} 
	\caption{ShieldScatter integrates with existing security protocols in the upper layers to enable device authentication and defend against attackers.}
	\label{fig:security}
\end{figure*}
\section{Security Analysis}\label{sec:security}
We finally integrate ShieldScatter with existing security protocols in the upper layers to enable device authentication and defend against active attacks. Sitting between upper-layer security protocols and PHY signal processing, ShieldScatter conforms to reasoning analogous to existing security protocols but differs in that ShieldScatter takes into account the propagation signatures to secure IoT devices.

\textbf{Deauthentication deadlock mitigation.} There are various ways to launch DoS attacks. A typical type of DoS attacks takes the vulnerability before a secure link has been established. As shown in Fig.~\ref{fig:security}(a), we consider that an authentication handshake is in progress. During the authentication handshakes, to deauthenticate the establishment, an attacker can inject an unauthorized deauthentication notification after receiving an acknowledgement (ACK) from the AP, which accordingly leads to a protocol deadlock.

To defend against the deauthentication deadlock, ShieldScatter adds an additional propagation signature processing at the AP with slight protocol changes. Specifically, ShieldScatter controls the message transmission of handshake within coherence time. Then, upon hearing the deauthentication command, ShieldScatter can compare the similarity between the deauthentication command and the following \textit{Message} with the operations mentioned in Section~\ref{sec:system}. If the one-class SVM identifies that the deauthentication command is from an attacker, the AP will drop this data frame in the upper layer. Accordingly, ShieldScatter can easily defend against such attacks.

\textbf{Jamming and replay attack mitigation.} An attacker can transmit jamming and replay attack by being equipped with multiple antennas. A multi-antenna attacker can jam the association packets reception with one directional antenna and records the packet with another directional antenna. In order to avoid jamming itself, it can use cables to test and separate the directional antennas at a proper distance before jamming the IoT device. Then, the attacker replays the recorded packets to the legitimate IoT device.

As illustrated in Fig.~\ref{fig:security}(b), we also take the handshake process and deauthentication deadlock into account. During this process, an attacker first injects an unauthenticated deauthentication notification (e.g., DoS) after receiving the ACK from the AP. When detecting the following \textit{Message 1}, the attacker jams this reception at the AP with one directional antenna, while at the same time records \textit{Message 1} with another directional antenna. The attacker then replays the recorded message to the AP. Thus, both of deauthentication command and the message are from the attacker and the multi-path signatures will be the same. However, when the attacker jams the reception of \textit{Message 1}, the multi-path signatures at each tag during jamming are the superposition of the legitimate device and attacker. This will lead to a large difference in the energy for each tag. Thus, ShieldScatter can easily detect this difference and defend against attacks. 

\textbf{Authentication deadlock mitigation.} 
According to~\cite{xiong2013securearray,eian2011modeling}, when a legitimate device and an AP are sharing messages with each other, they easily suffer from an authentication deadlock caused by an active attacker. Specifically, as shown in Fig.~\ref{fig:security}(c),  during a data sharing process, after detecting the message from the IoT device, an active attacker would transmit an authentication request to the AP. Thus, the AP will be misled to a wrong state and it will reject the message from the legitimate IoT device, which accordingly leads to an authentication deadlock.

To defend against the authentication deadlock, ShieldScatter needs only a slight protocol change at the AP end. Specifically, ShieldScatter controls the duration between the message transmission and authentication request within the coherence time. Then, upon hearing the authentication request, ShieldScatter can compare the backscatter signals' similarity between the authentication message (e.g., \textit{Request}) and the previous message (e.g., \textit{Message i}) with the operations mentioned in Section~\ref{sec:system}. If the system identifies that the authentication request is from an attacker, the AP will remove this data frame in the upper layer, and thus, ShieldScatter can successfully defend against this attack.

\section{Implementation}\label{imple}
\begin{figure}[t]
	\center
	\includegraphics[width=2.0in]{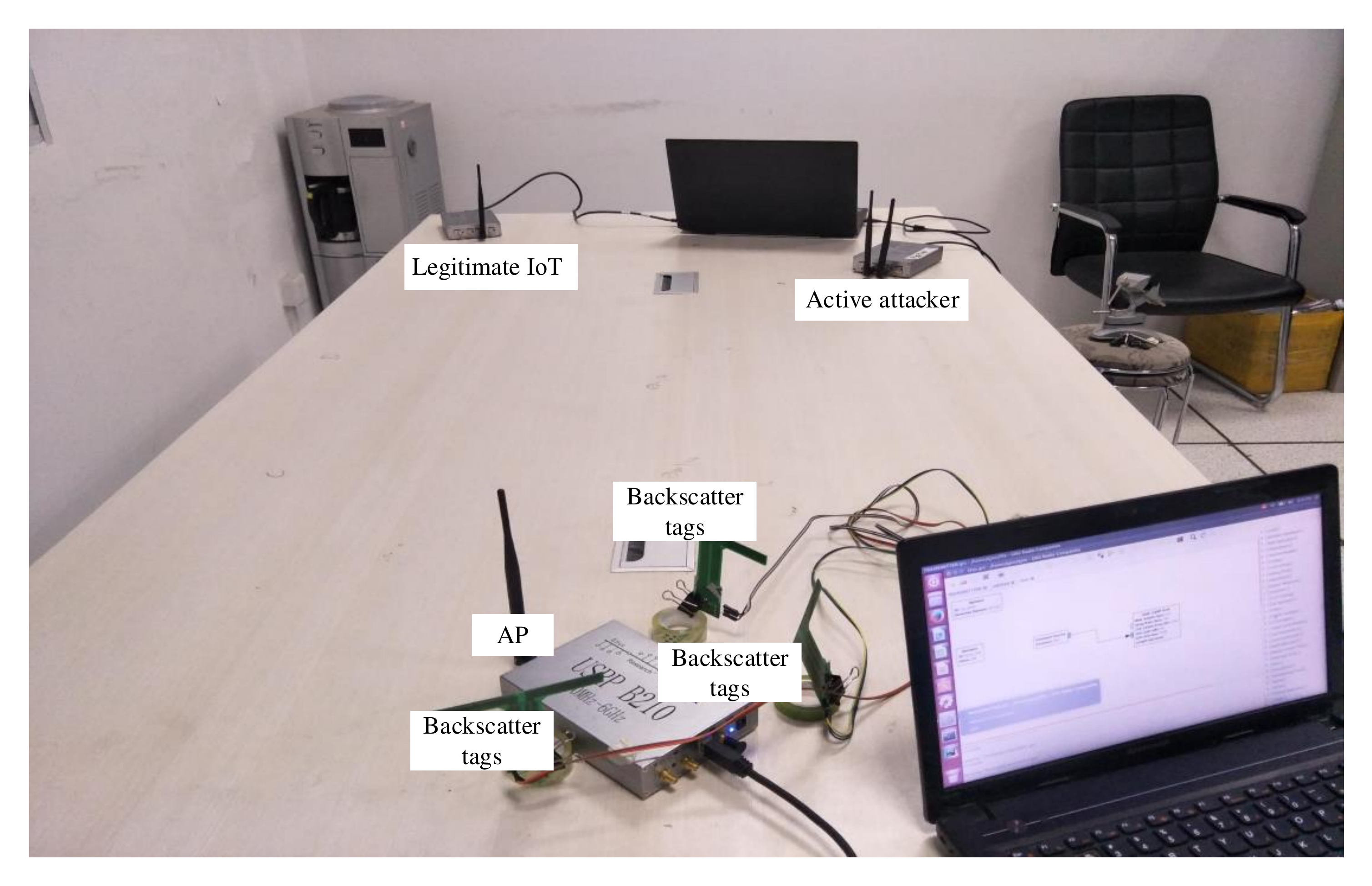} 
	\caption{We employ two USRPs to act as the AP and legitimate IoT device at a distance of 2.5~m. At the same time, several tags are deployed around the AP to create multi-path signatures. Besides, another two-antenna USRP is used to act as active attacker. }
	\label{fig:imple}\vspace{-0.3cm}
\end{figure}

\textbf{Prototype.} As shown in Fig.~\ref{fig:imple}, the prototype of ShieldScatter is implemented using multiple backscatter tags and three GNURadio/USRP B210 nodes. We tailor the antenna design to allow tags to work at 900 MHz which is a commonly used frequency for IoT devices. In our experiment, all tags transmit data with a bitrate of 10 kbps. Besides, one USRP node equipped with two antennas acts as an active attacker, who monitors RSS variations with one of the antennas while transmitting fake data using the other antenna. The other two USRP nodes are used as the legitimate IoT device and the AP, respectively and each of them contains only one antenna. In our experiment, we require the IoT device and the AP to complete the challenge-response protocol in 100~ms so as to ensure that the channel is stable.

\textbf{Tag deployment.} According to recent studies~\cite{tse2005fundamentals,pierson2018poster}, if two devices are placed at a distance less than half a wavelength, the signal propagation of these two devices will experience a similar channel state. Thus, if we place all the tags within a short distance, the backscatter signals from these tags will be highly correlative and indistinguishable. Accordingly, the distances among any tags should be larger than half of the wavelength. Besides, since ShieldScatter secures the devices by relying on the multi-path signatures of the tags, an attacker may have similar multi-path profiles with the legitimate IoT device when the two are placed in symmetric locations. Thus, in order to prevent such case, we deploy all the tags at various angles on a circle with a distance of 15~cm.

\textbf{Testing environment.} In our experiment, we evaluate the performance of ShieldScatter in both static and dynamic environments as shown in Fig.~\ref{fig:floorplan}. We first deploy two USPRs at a distance of 2.5~m. Specifically, we place the legitimate IoT device at different locations (e.g., the blue blocks)  and the AP at location B, respectively. Besides, another USRP that contains two antennas is deployed at different locations (e.g., the red dot in Fig.~\ref{fig:floorplan}) to act as an active attacker. We conduct our experiment in a static environment and collect the signals day and night. As for dynamic environment, two volunteers are asked to walk around when we perform the experiments.

In our experiment, in order to remove the effects of IoT temperature, weather and humidity, we collect and calculate the propagation signatures over one month for ShieldScatter in both static and dynamic environments. Then, ShieldScatter extracts the features to construct the profiles for all of these data samples. 577 of the propagation profiles are used to train and construct our one-class SVM model and the rest of the profiles are used to test the performance of ShieldScatter.

\textbf{Metrics.} We employ the following metrics to evaluate the performance of our system. \begin{itemize}
	\item \textbf{True positive rate.} True positive (TP) rate is the ratio of the number of samples that are correctly identified as being from the legitimate IoT device to the total number of positive samples.
	\item \textbf{False positive rate.} False positive (FP) rate is the ratio of the number of the samples that are falsely recognized as coming from the legitimate IoT device to the total number of negative samples.
\end{itemize}

\section{Evaluation}\label{sec:eva}
\subsection{Parameter Determination}\label{sec:parameter}
\begin{figure}[t]
	\center
	\includegraphics[width=2.0in]{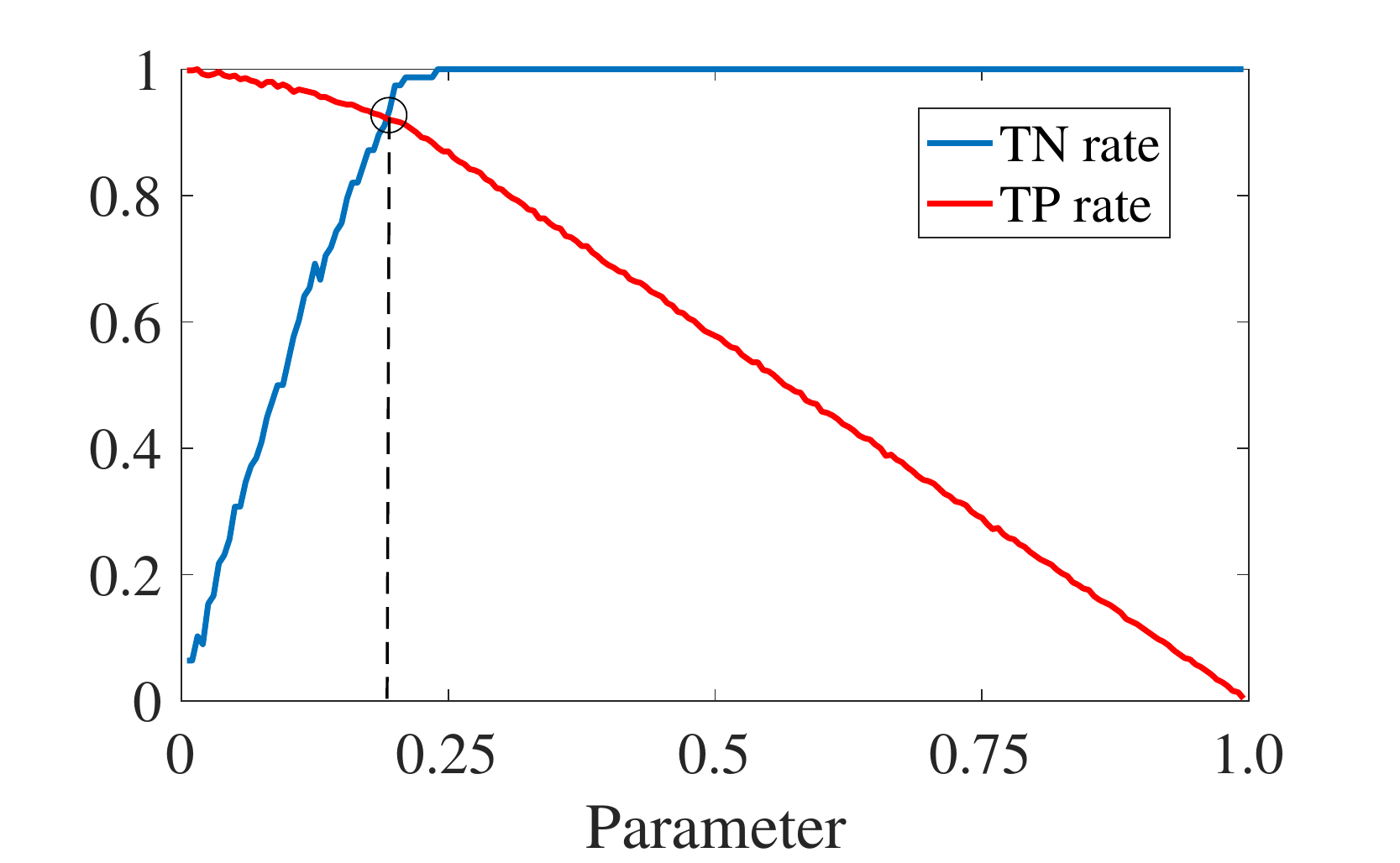} 
	\caption{The performance with respect to the varying parameter $v$. }
	\label{fig:parameter}\vspace{-0.3cm}
\end{figure}

 As mentioned before, $v$ is a significant parameter to constrain the boundary of outliers and support vectors, and the range of $v$ is $v \in (0,1]$. In order to determine the parameter $v$, we exploit 500 groups of positive data samples and 77 groups of negative samples for input to training the one-class model varying the parameter $v$ from 0 to 1. As shown in Fig.~\ref{fig:parameter}, it is obvious that the accuracy of correctly detecting the legitimate IoT device decreases with the parameter $v$. However, the accuracy in detecting the active attackers increases when the parameter $v$ grows. That is because when parameter $v$ increases, the boundary between the positive samples and negative samples tightens. Consequentially, the outliers (i.e., the attacker) are excluded from the positive samples. However, the larger the parameter increases, the smaller the boundary becomes. In that case, some positive samples are recognized as the outliers and excluded from the positive samples. Thus, in order to determine the optimal $v$ for the one-class SVM model, we make a tradeoff between them. As shown in Fig.~\ref{fig:parameter}, we select parameter $v$ at the intersection point between these two curves. Accordingly, we can achieve the accuracy of 93.7\% for legitimate IoT device and active attacker detection when the parameter $v$ is set to 0.16.
\begin{figure}[t]
	\center
	\includegraphics[width=2.3in]{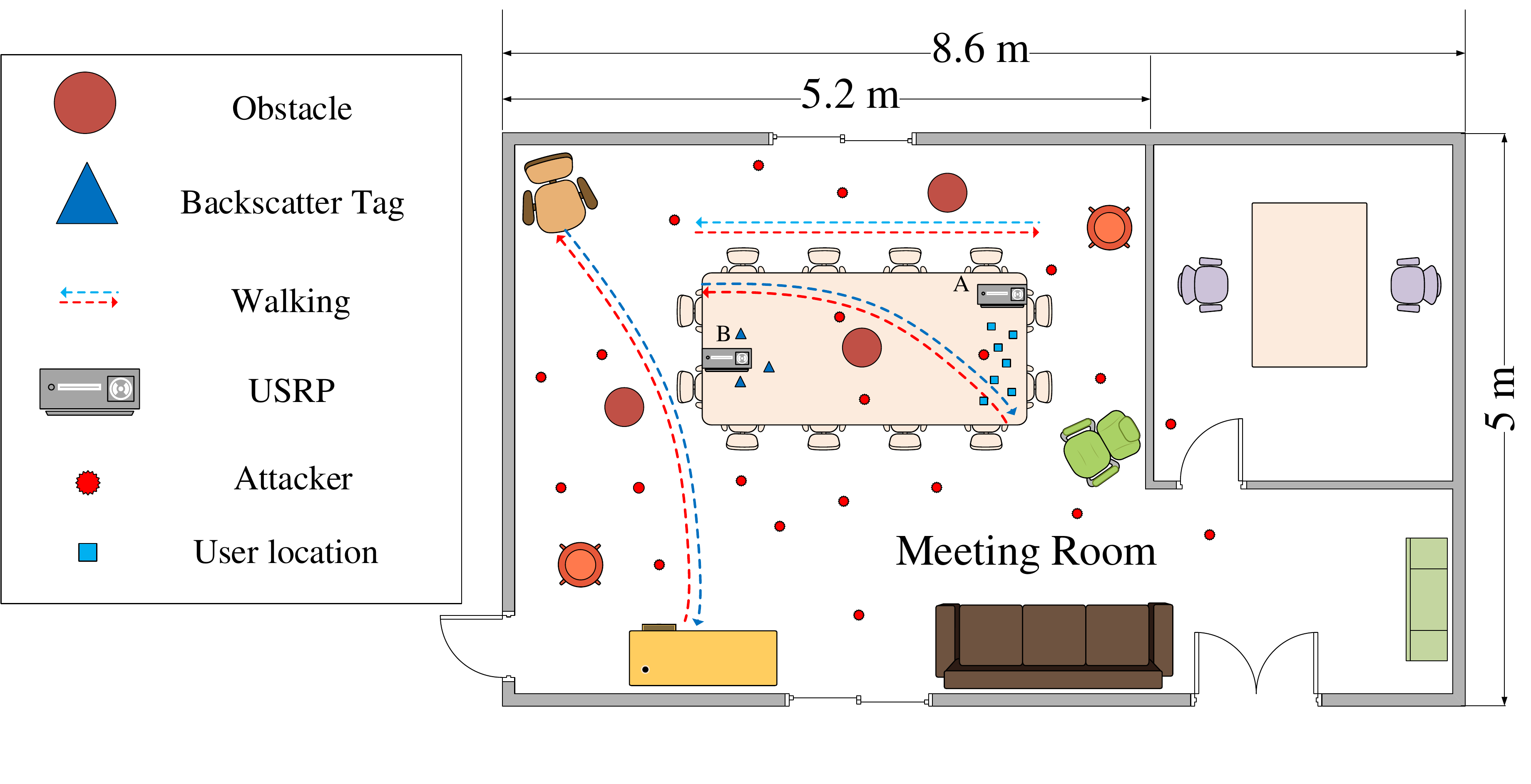} 
	\caption{Floor plan of our evaluation environment. A and B represent the positions of legitimate IoT device and AP, respectively. }
	\label{fig:floorplan}\vspace{-0.3cm}
\end{figure}

\begin{figure}[t]
	\center
	\includegraphics[width=2.0in]{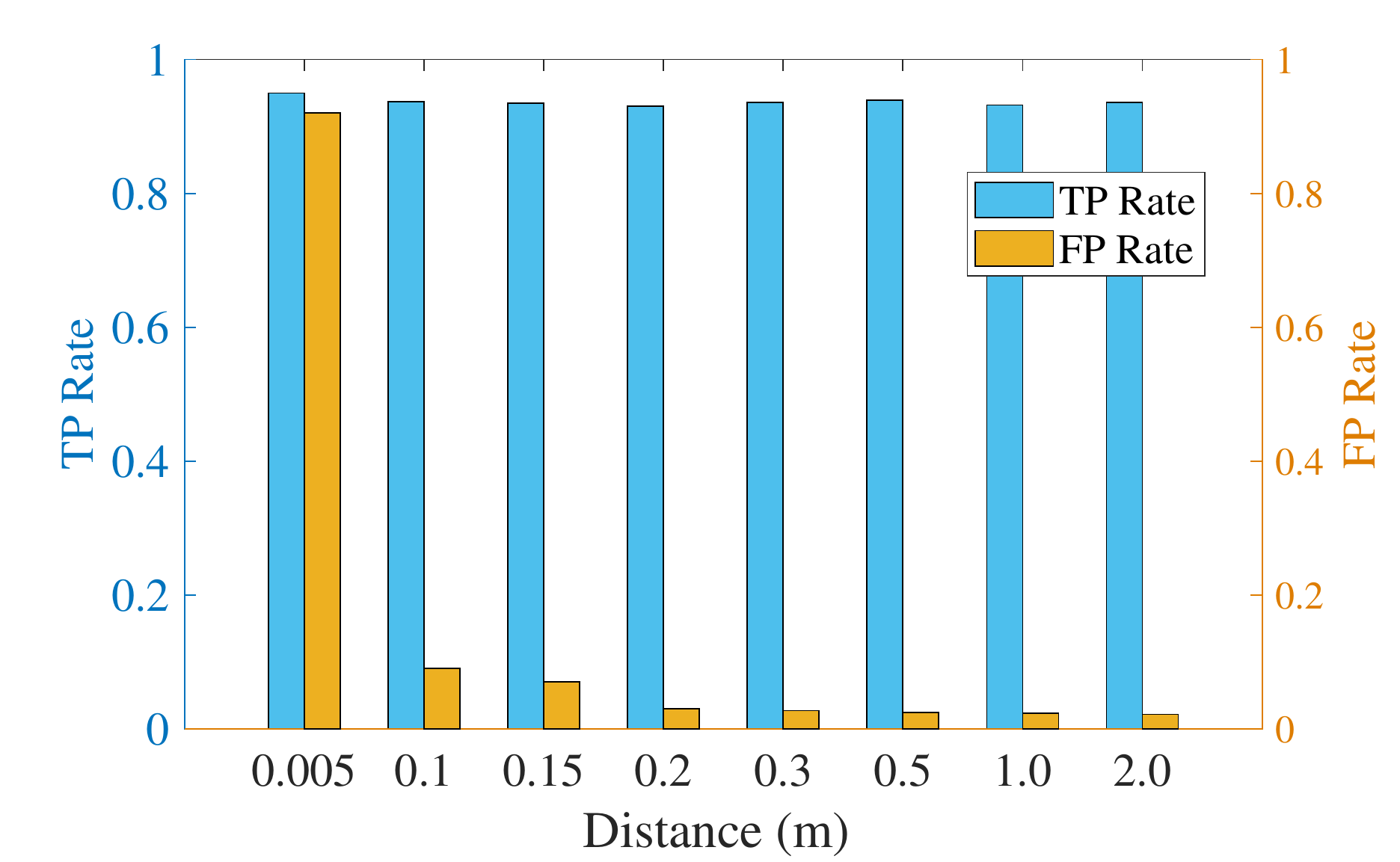} 
	\caption{The performance with respect to the varying distances between the attacker and legitimate IoT device in a static environment. }
	\label{fig:static_distance}\vspace{-0.3cm}
\end{figure}

\begin{figure}[t]
	\center
	\includegraphics[width=2.1in]{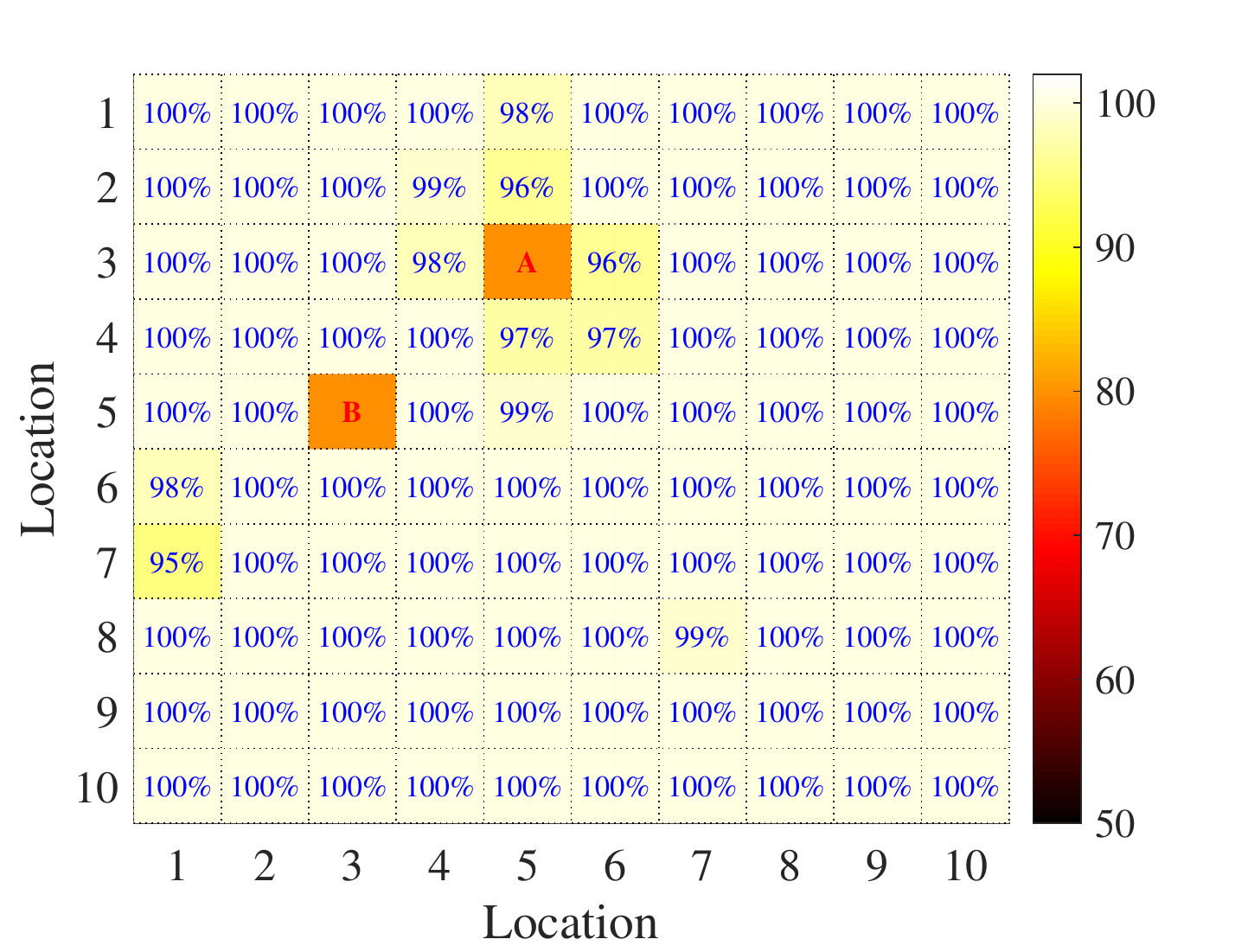} 
	\caption{We divide the room into 100 grids and evaluate the performance in each location, where A and B is the legitimate IoT device and AP, and we also attached three tags around the AP at a distance of 15cm.}
	\label{fig:locations}\vspace{-0.3cm}
\end{figure}

\subsection{Basic Attack Defense}\label{sec:basic}
We first evaluate the performance of ShieldScatter in the static environment, where we keep the legitimate IoT device, the AP, and the attacker static. Then, we evaluate ShieldScatter in the dynamic environment. Finally, we compare with a simple and commonly used correlation approach to show the advantage of our system.

\subsubsection{Static environment}\label{sec:static}

\textbf{All the devices are in static state.} Usually, the attacker will be far away from the legitimate IoT device. However, if the attacker is small enough and has the ability to get close to the system, it will be a challenge for the IoT devices.Thus, in order to defend against the attacker that is close to the legitimate device, we first validate the performance when the attacker is placed very close to the system, where we deploy it at a distance of 0.5cm from the IoT device. Then, we evaluate the performance of our system by changing the distance between the legitimate device and the active attacker ranging from 10~cm to 2~m.

As shown in Fig.~\ref{fig:static_distance}, the attacker can successfully spoof the  AP when it is placed in a very close location to the IoT device. Besides, if the legitimate device is close to the attacker, especially when the distances are lower than 15~cm, the FP rate increases dramatically. However, when we move the attacker away from the legitimate device, ShieldScatter can achieve an average TP rate of 93.6\% and FP rate of 3\%. That is because if the attacker is close enough to the legitimate device, their multipath profiles caused by the backscatter tags are extremely similar. In our experiment, even though the attacker is located only 15-20cm away from the legitimate device, we can still mitigate 95\% attacking attempts. We believe that an attacker would not be installed in such a close distance and it is acceptable for the daily smart home loT devices. Otherwise, we can easily find it if the attacker is placed as close as 15-20cm to the legitimate IoT device.

 As for the other locations of the room, we divide it into 100 small grids ($ 10\times10$), then we extract 100 samples to test the results in each location. According to Fig.~\ref{fig:locations}, we can almost detect all the attackers in most of the locations and the attacker cannot always successfully attack the system in a fixed location. The reason is that the channel state will vary in different time even though the attacker is placed in a fixed position. Therefore, the radio propagation will be significantly different from the legitimate IoT device. However, we consider the security problem happens when a legitimate device is pairing or sharing data with the AP. In this case, this process is within the coherence time and the channel would stay stable. As a result, the similarity of these two signals from the legitimate IoT devices will not be significantly changed within such a short duration.

\textbf{Experiment in various environments.} In this section, we also evaluate the performance in various environments, including the laboratory, the meeting room, and the corridor. In particular, the lab environment exists many reflectors, like the desks and a steel closet. In contrast, the meeting room is empty, with fewer obstacles, and the corridor is narrow and long. Then, we collect the backscatter features to train and test in all these three different environments. In addition, we also employ the training model in the laboratory to evaluate the performance in the other two environments.

\begin{figure}[t]
	\center
	\includegraphics[width=2.4in]{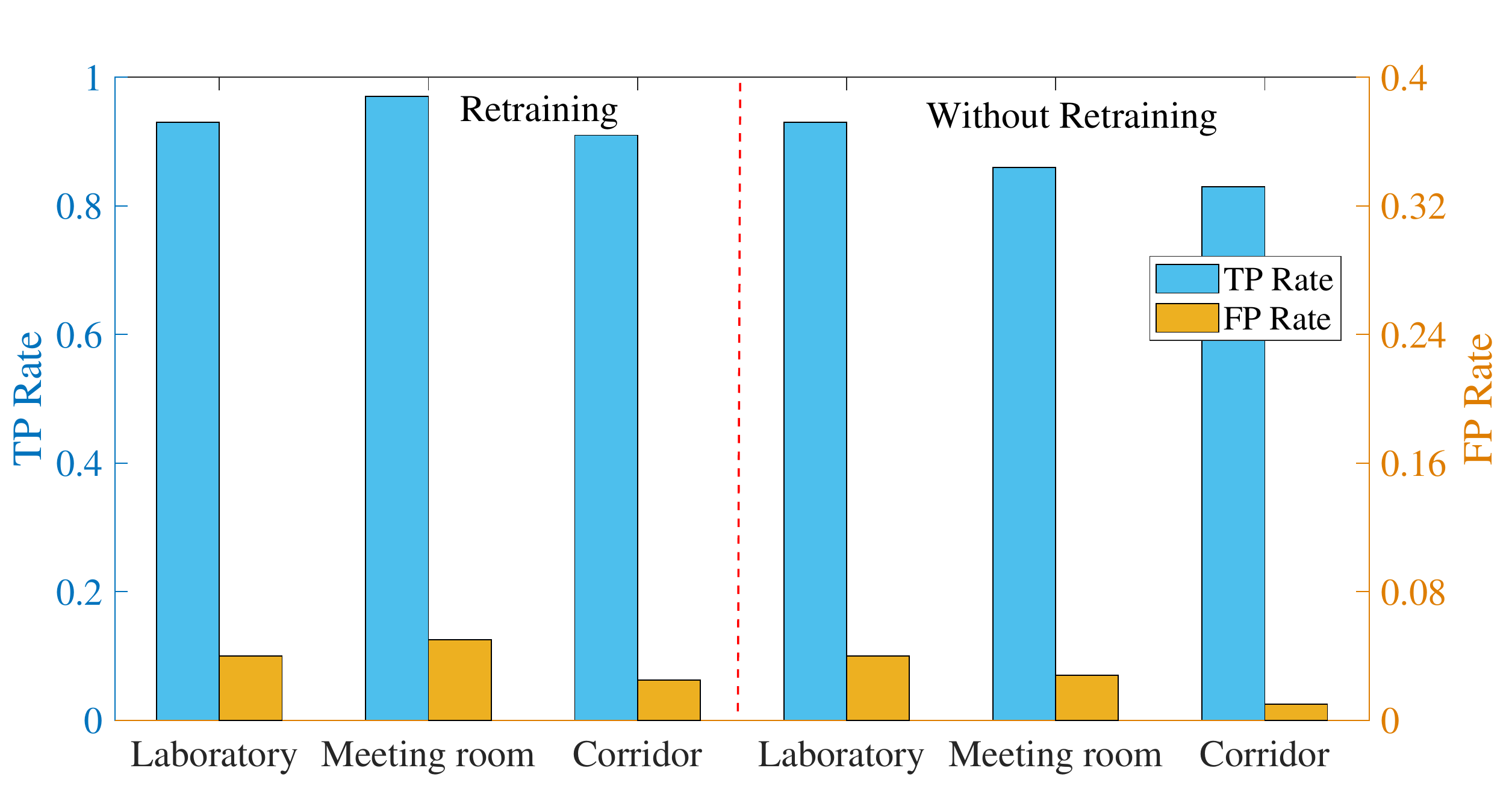} 
	\caption{We conduct experiment to evaluate the performance in various environments, including the laboratory, meeting room and corridor. }
	\label{fig:rooms}\vspace{-0.3cm}
\end{figure}

As presented in Fig.~\ref{fig:rooms}, when we train and test the results in these three different environments, we can all achieve the average TP rate larger than 93\% and FP rate lower than 5\%. In contrast, if without retraining, we achieve the average FP rates of 2.8\% and 2.6\% in the meeting room and the corridor. That is because the channel and location changes lead to more differences between the IoT device and attacker. However, since the new environment has significantly changed the channel, even though the two signals are from the same device within the coherence time, the differences created by the channel change would be still increased. Accordingly, we have lower TP rates of 86\% and 83\%, respectively. Therefore, if the system moves to a new environment, we can retrain the model or authenticate the IoT device many times with a voting scheme to maintain a better TP rate. 

\textbf{Number of backscatter tags.}
In our experiment, the number of the backscatter tags used to be deployed around the AP and construct multi-path signatures is an important factor for the radio propagation. Thus, we then evaluate the performance when using a different number of backscatter tags attached around the AP. 

\begin{figure}[t]
	\center
	\includegraphics[width=2.0in]{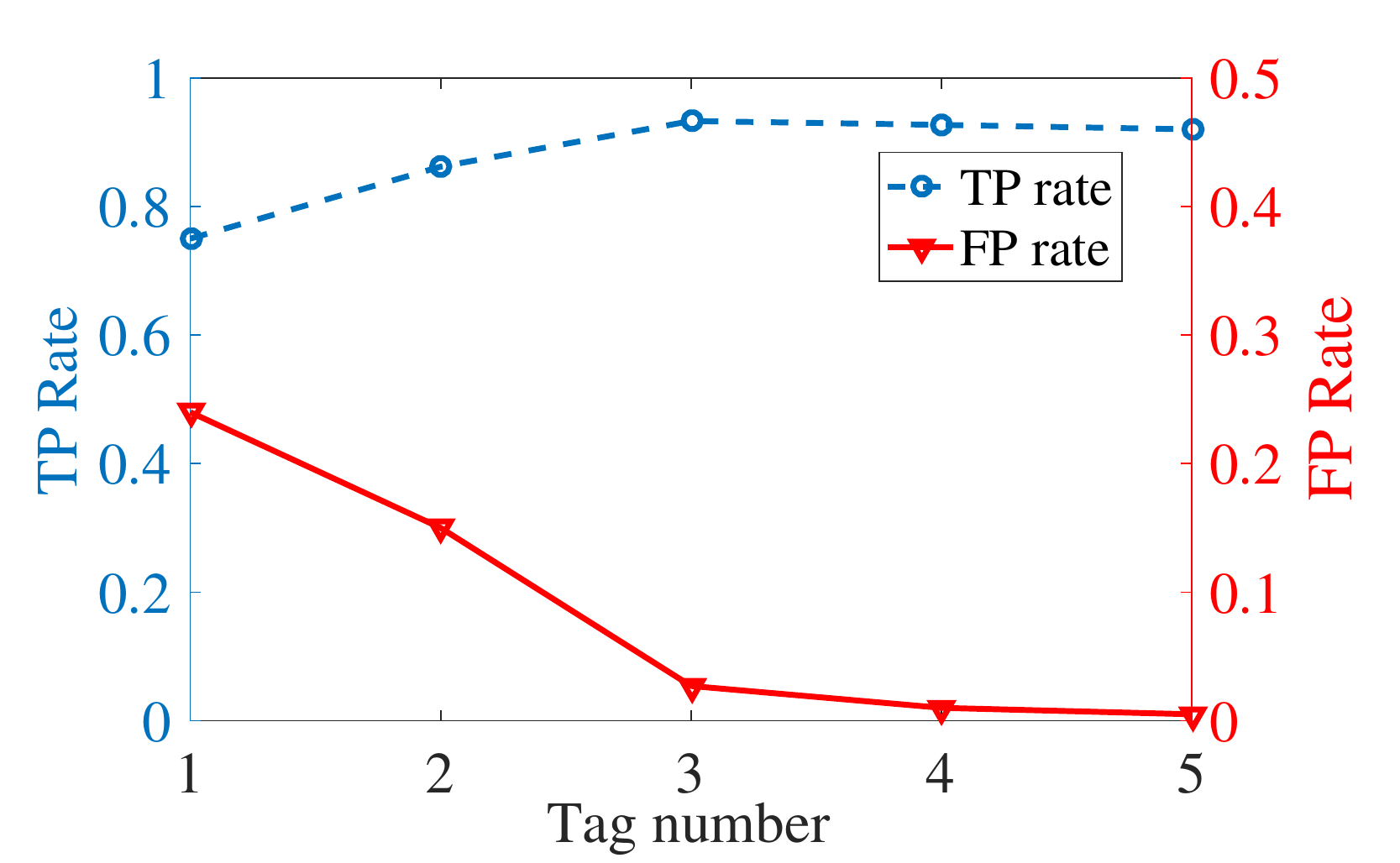} 
	\caption{The performance with respect to the varying number of backscatter tags. }
	\label{fig:tags_num}\vspace{-0.3cm}
\end{figure}

As shown in Fig.~\ref{fig:tags_num}, we can observe that if we just attach one or two backscatter tags on the AP, ShieldScatter can achieve an average FP rate higher than 10\%. It is because the active attacker can easily calculate the power between the legitimate device and the AP. Then the attacker can carefully select the transmitting power and locations to attack. However, if we deploy three tags, it is harder for the attacker to keep the received power for each tag as the legitimate IoT device. Hence, we can achieve an average TP rate as high as 93.7\% and FP rate lower than 3\%. However, if we exploit too many tags (e.g., 4-5 tags), it leads to strict constraints on the received signal, and accordingly achieve lower FP and lower TP rates. Therefore, three backscatter tags are an appropriate choice for our system.

\begin{figure}[t]
	\center
	\includegraphics[width=2.0in]{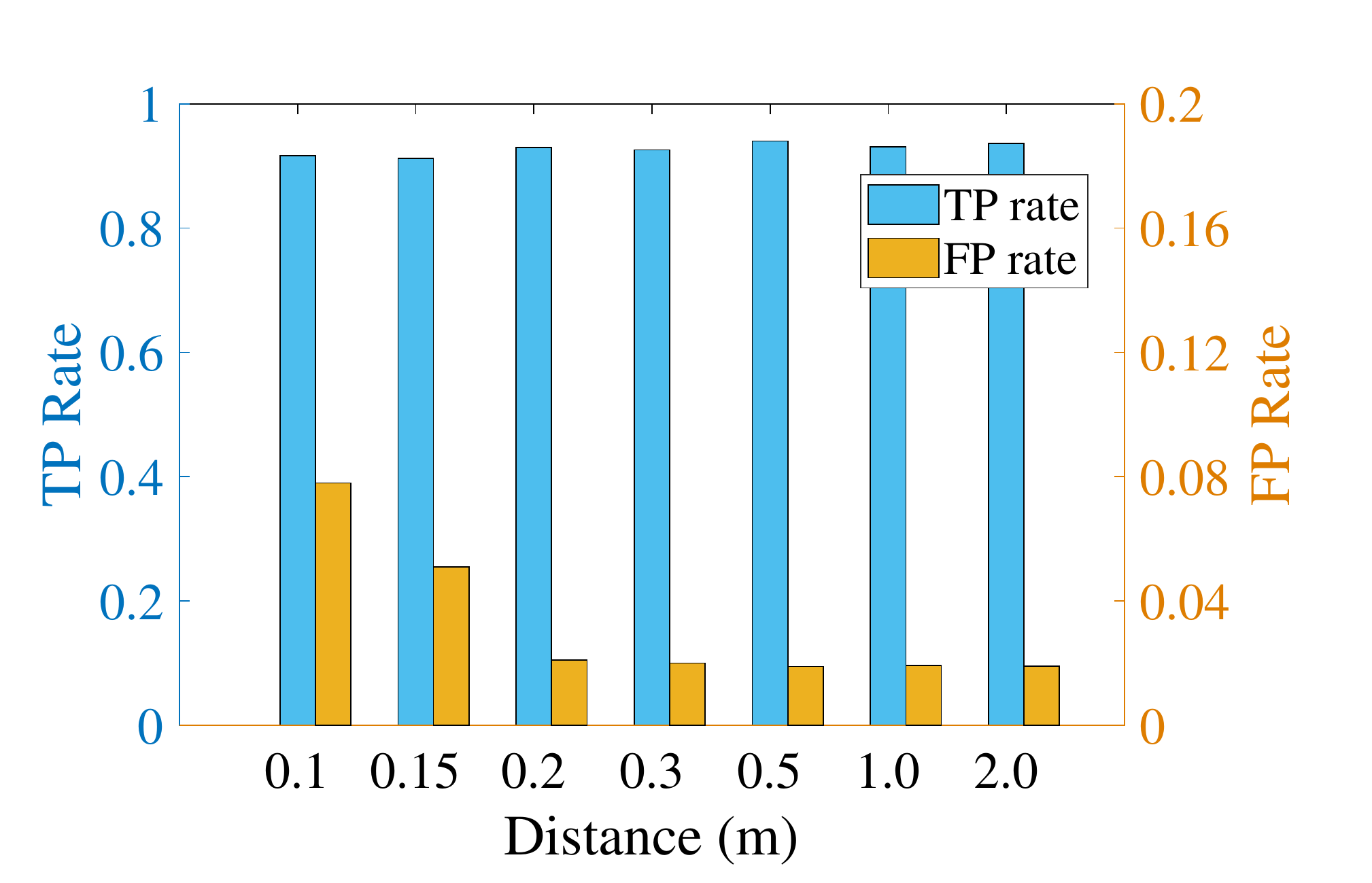}\vspace{-0.2cm}
	\caption{The performance with respect to the varying distances between the attacker and legitimate IoT device in a dynamic environment. }
	\label{fig:mobile_distance}\vspace{-0.3cm}
\end{figure}

\subsubsection{ Dynamic environment }\label{sec:mobile}

\textbf{People walking around.} ShieldScatter considers another practical environment for the daily smart home devices, that is, the scenario where some people are walking around.  As shown in Fig.~\ref{fig:floorplan}, in order to emulate the dynamic environment, two volunteers are asked to walk around the devices, approach the device, go across the channel and carry out daily activities. Then,  we evaluate the performance with respect to different distances between the legitimate IoT device and the attacker.

As shown in Fig.~\ref{fig:mobile_distance}, compared with the results in a static environment, when the environment is dynamic, the TP and FP rates of ShieldScatter have a slight fluctuation caused by wireless channel and environmental noise. However, since the challenge-response process is completed within the coherence time and we have also exploited the filter to remove the noise caused by the dynamic effects, ShieldScatter can still achieve an average TP rate higher than 91\%, which is acceptable for smart home IoT devices. Besides, ShieldScatetr achieves a better average FP rate lower than 1.9\%, when the distance is larger than 20 cm. That is because the channel fluctuation makes it more difficult for the attacker to emulate the signal for each tag. Thus, our system can remain reliable even in a dynamic environment.

\textbf{Slight movement of the legitimate IoT device.} ShieldScatter takes into account the case where the smart home devices are moved slightly. Specifically, as shown in Fig.~\ref{fig:floorplan}, the legitimate IoT device is first placed at location A, and then it is moved to a different place as the blue blocks. Then, we evaluate the performance of our system with respect to the distance between location A and the device.

\begin{figure}[t]
	\center
	\includegraphics[width=2.0in]{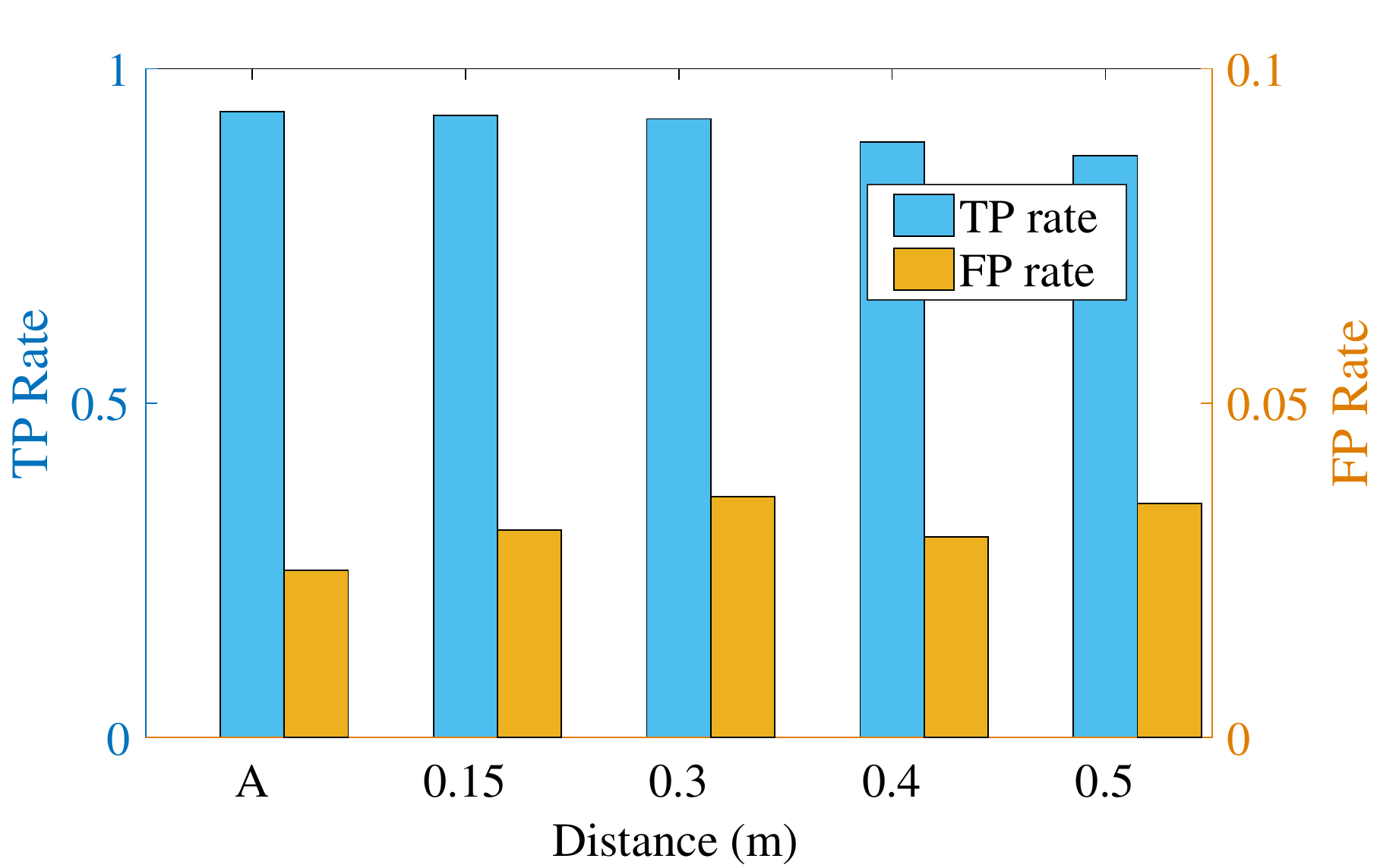} 
	\caption{The performance when the legitimate IoT device is slightly moved. }
	\label{fig:movement}\vspace{-0.3cm}
\end{figure}

As shown in Fig.~\ref{fig:movement}, we can achieve an average FP rate lower than 3\% when the device is moved to different locations. On the other hand, when the legitimate IoT device is placed in location A, we can achieve a TP rate of 93\%. Then, if we move the device a short distance (e.g., within 30 cm), the TP rate remains stable. When the device has moved a long distance, the TP rate would decrease. However, we can still achieve a TP rate larger than 87\% even though the movement distance is 50 cm. Therefore, ShieldScatter can significantly secure the IoT device for the smart home.

\begin{figure}[t]
	\center
	\includegraphics[width=2.1in]{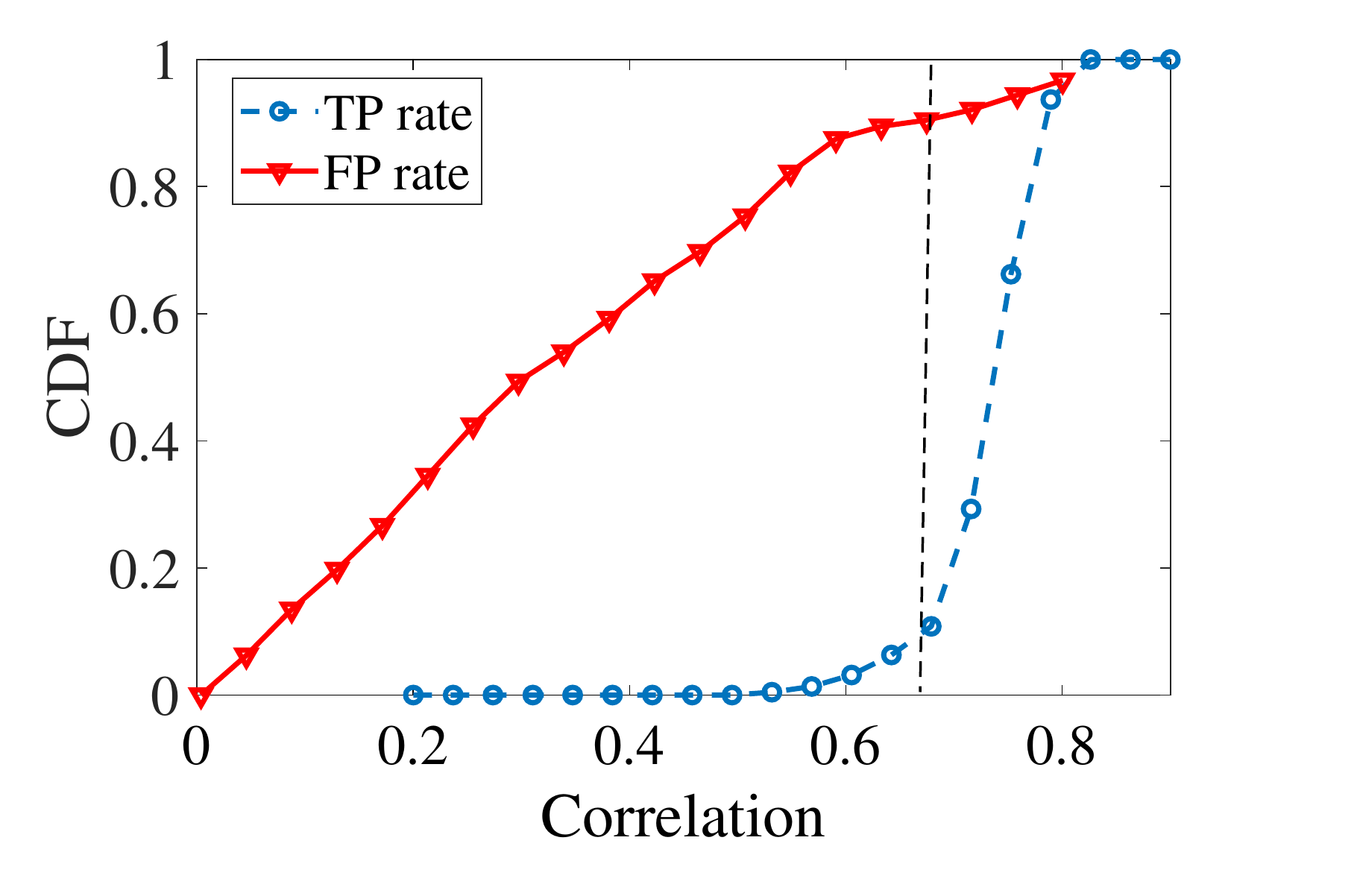} 
	\caption{The performance when we exploit a correlation method.}
	\label{fig:correlation}\vspace{-0.3cm}
\end{figure}		
\subsubsection{ Correlation approach }\label{sec:compare}
In this section, we evaluate the performance when using a simple correlation approach in a static environment. Specifically, we first extract the backscatter signals from the received raw data. Next, we calculate the correlation between these two signals. Finally, by calculating the cumulative distribution function (CDF), we can select an appropriate threshold to decide whether the received signals are from the legitimate IoT device.

We calculate the CDF with respect to the TP rate and FP rate as shown in Fig.~\ref{fig:correlation}. According to the results, we find that, if the correlation threshold is large, we can achieve lower TP rate and FP rates. On the contrary, if we select a small one, we will achieve a better TP rate but a larger FP rate. Thus, we can make a tradeoff to decide the threshold where we select the correlation threshold to be 0.6789. Then, based on this threshold, our system can achieve a TP rate of 89\% and FP rate of 9.5\%. Compared with the results using our algorithm (with a TP rate of 93.7\% and FP rate of 3\%), ShieldScatter is more reliable to secure the devices.

\subsection{ Advanced Attack Defense}\label{sec:powerful}
In this section, we evaluate the performance when defending against an advanced attack with two schemes.

\textbf{Tag-random scheme.} In this experiment, we evaluate the performance when the legitimate IoT device is attacked by a powerful attacker. In particular, to impersonate a powerful attacker, we first measure the power and signals reflected by each tag as well as the signals directly transmitted from the IoT to the AP. Then, we tune an antenna to focus at the AP and directly launch this responding power and signals. Accordingly, we can guarantee that the signals from this attacker are similar to those from the legitimate device. To validate the powerful attacker, we first evaluate the performance when the attacker transmit a well-designed signal to attack the system at a distance of 2.5m from the IoT device. Then, we evaluate the performance when the attacker is placed at various distances from the AP.
\begin{figure}[t]
	\center
	\includegraphics[width=1.8in]{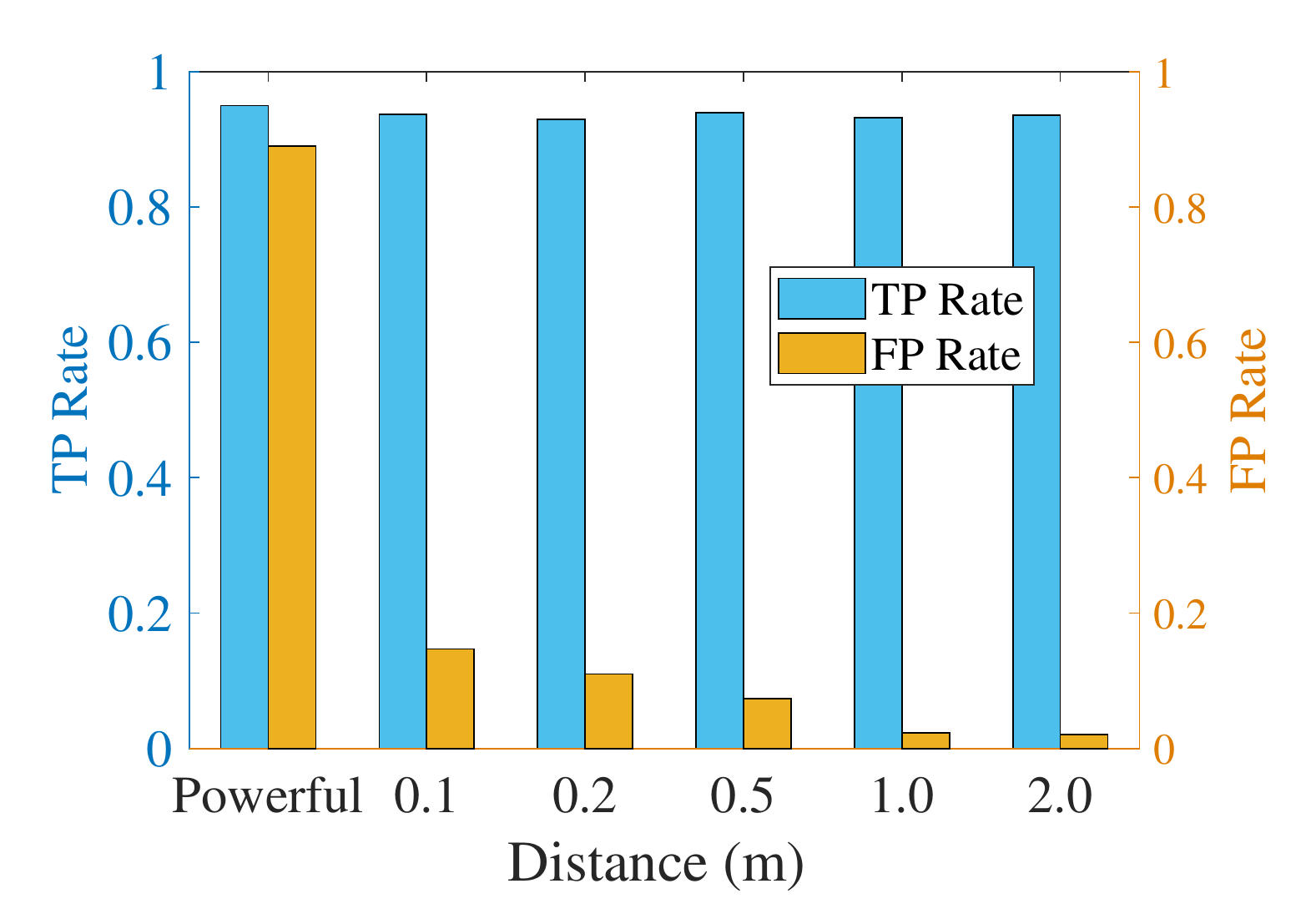} 
	\caption{The performance when we design a tag-random scheme to defend against a powerful attacker.}
	\label{fig:pow}\vspace{-0.3cm}
\end{figure}

As presented in Fig.~\ref{fig:pow}, when the powerful attacker attack the system with a well-designed signal, it can achieve an FP rate of 89\%. Besides, if an attacker is close to the AP, we can achieve a TP rate of 93\%  and FP rate of 14\%. Otherwise, we will achieve a better performance with an FP rate as low as 7\%, which is reasonable. First, because of the random reflection orders, the attacker cannot predict the orders of our tags. Therefore, the attacker is unable to estimate the real channel impulse response and fails to spoof the AP. Second, owing to the imperfect design of antennas, it brings extra noise to the system, making it more difficult for the attacker to design the actual signals to the AP when the attacker is far from the receiver. Thus, our proposed scheme can successfully defend against a powerful attacker.

\textbf{Multiple receivers combination scheme.} We evaluate the performance by combining multiple receivers. In particular, we first deploy only one USRP as the AP and attach three tags around this receiver to reflect the signals. In addition, we also place several USRPs around the receiver to impersonate other APs. At the same time, we also use a USRP as a powerful attacker placed at a distance of 20cm from the legitimate IoT device. We evaluate the performance with only one receiver and also test the results by increasing the number of APs. 

\begin{figure}[t]
	\center
	\includegraphics[width=2.0in]{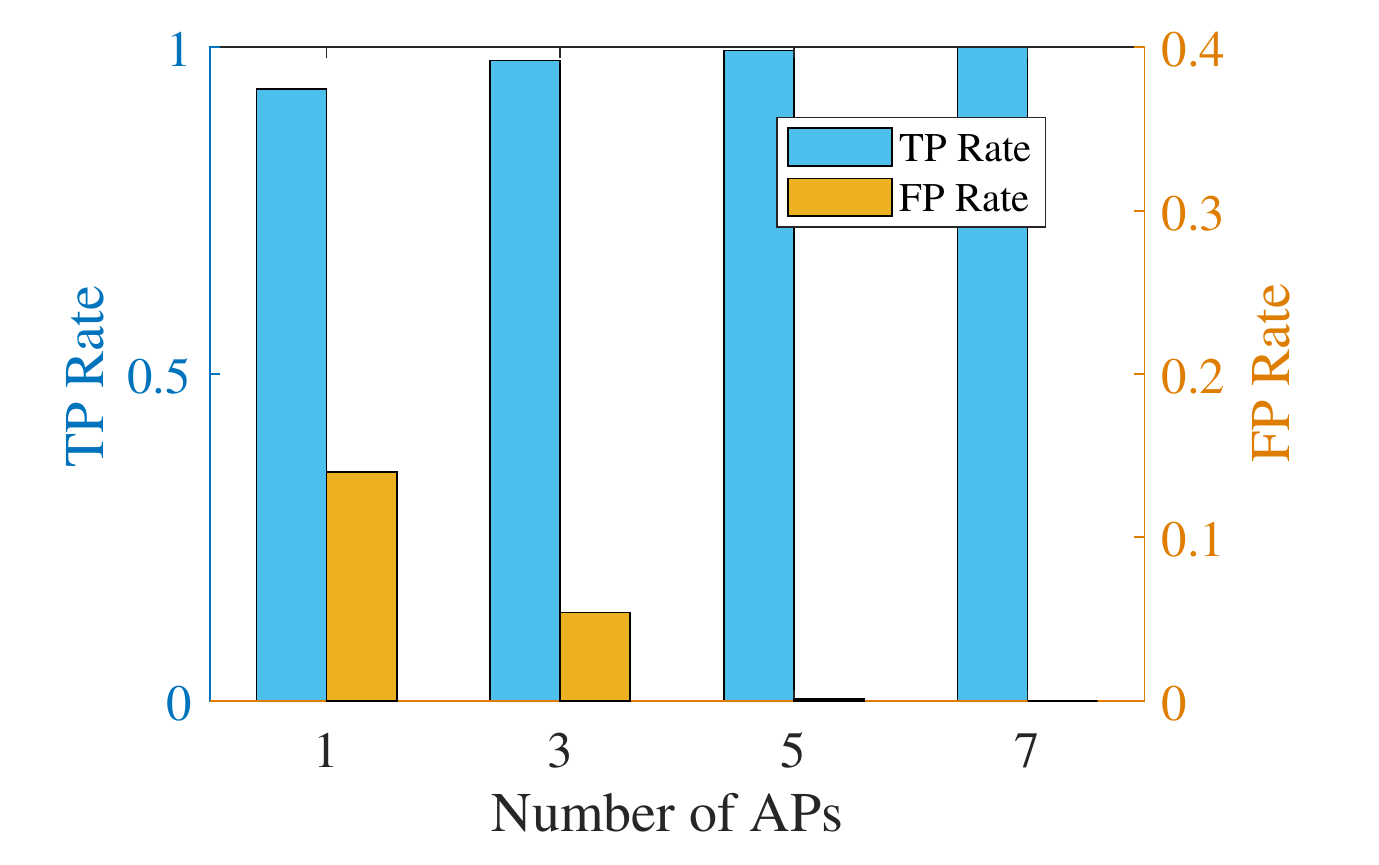} 
	\caption{The performance when we combine different number of APs.}
	\label{fig:combination}\vspace{-0.3cm}
\end{figure}
\begin{figure}[t]
	\center
	\includegraphics[width=2.0in]{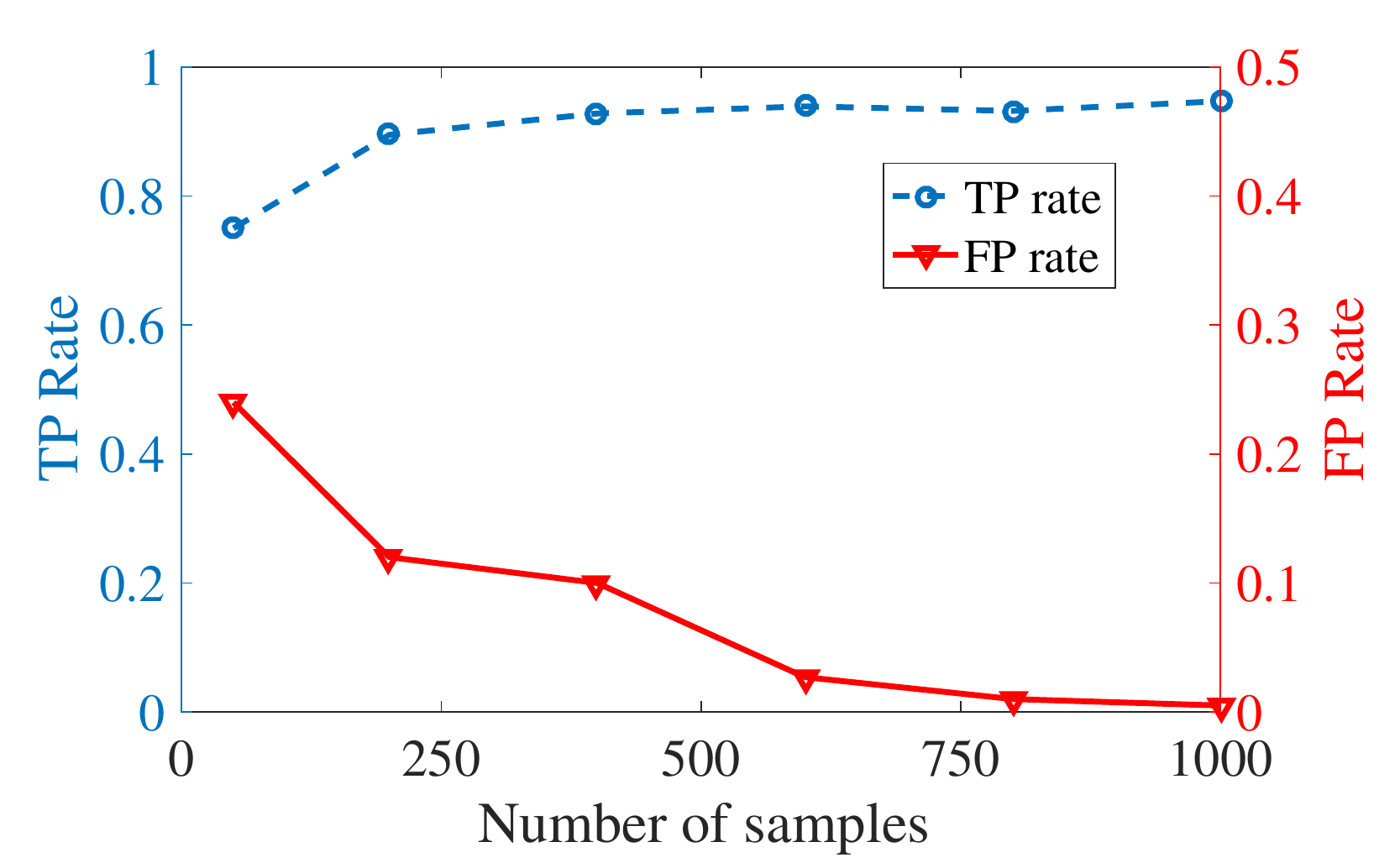} 
	\caption{The performance with respect to the varying number of data samples to train the model. }
	\label{fig:samples_num}\vspace{-0.3cm}
\end{figure}

As shown in Fig.~\ref{fig:combination}, when we use only one AP, if the attacker is place at a distance of 20 cm from the legitimate IoT device, we can achieve a TP rate of 93\% and FP rate of 14\%, respectively. However, if we increase the number of APs and identify the attacker with a voting scheme, according to the results, we can achieve a TP rate of 98\% and a FP rate of 5\% when using 3 APs, and almost recognize the legitimate IoT device and detect the attacker with 100\% accuracy when combining 5 or 7 APs. The results show that our system can successfully detect the attacker by combining multiple APs.

\subsection{ Impact of SVM Model}\label{sec:impact}

\textbf{Training size.}
In our experiment, ShieldScatter needs to exploit a training set to train a one-class SVM classifier. Then, we evaluate the performance of the trained classifier on the testing samples. Thus, it is necessary to select a appropriately sized training set to train and construct the classifier for ShieldScatter. In particular, we train our model with respect to a different number of profiles ranging from 50 to 1,000 samples.

As described in Fig.~\ref{fig:samples_num}, when the data samples used to train the one-class SVM model are less than 200, ShieldScatter can achieve an average TP rate lower than 90\%. However, when the size of the training set is larger than 600, we can achieve a relatively reliable TP rate of 93.6\% and FP rate lower than 3\%. Consequently, we leverage 577 samples to train and construct our profile.

\textbf{The ratio of positive to negative samples.}
Based on the priori knowledge about one-class SVM, the ratio of positive to negative samples that used to train the model is an important factor. Thus, we evaluate the performance combined with different ratio of between them. According to~\cite{scholkopf2001estimating}, the number of the positive samples must be much larger than that from the attacker. Therefore, a noticeable constraint for one-class SVM model is that the ratio of positive to negative samples should be very large. Thus, we study the performance of ShieldScatter with respect to the low ratio of positive to negative samples ranging from 0.05 to 0.5.

\begin{figure}[t]
	\center
	\includegraphics[width=2.0in]{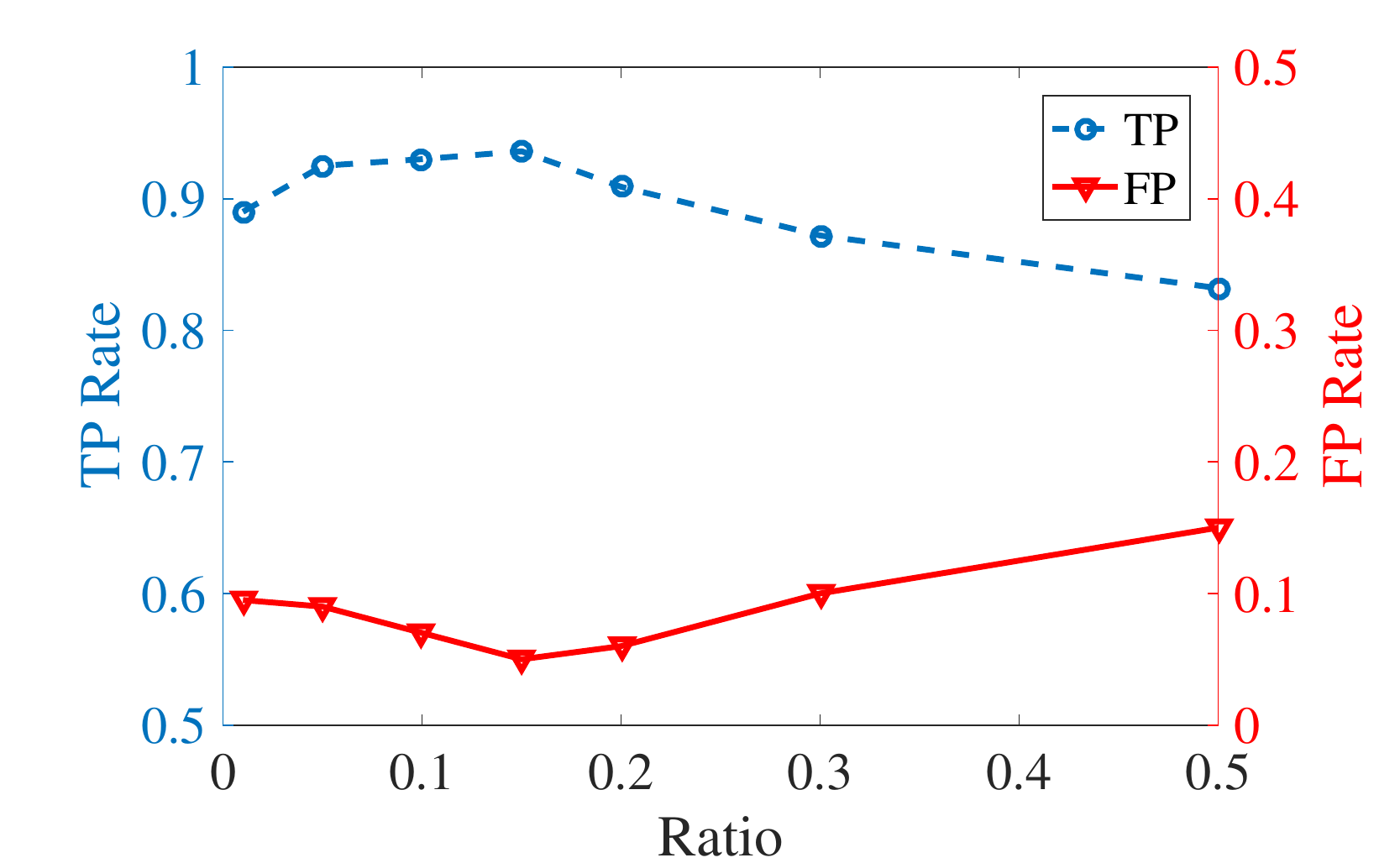} 
	\caption{The performance with respect to the varying ratios of positive to negative samples.}
	\label{fig:ratio}\vspace{-0.3cm}
\end{figure}

As presented in Fig.~\ref{fig:ratio}, when the size of the negative samples is too small, for example, the ratio of positive to negative samples is lower than 0.05, ShieldScatter achieves an average TP rate lower than 90\% and FP rate larger than 10\%. This is because if the size of negative sample is too small, the suspicious signals on the boundary are circled in legitimate IoT device but the positive samples are moved out. Conversely, if the size of the negative samples is too large, the model is unable to distinguish the positive and negative samples, which will lead to lower detection accuracy for the model. Therefore, ShieldScatter selects the ratio of positive to negative samples as low as 0.154 for the model training.

\textbf{Training without attacker samples.} 
In some cases, we may not have any priori knowledge of the attacker samples. Thus, we also explore the performance in various environments without using any attacker samples. In particular, we train the one-class SVM model and tune the parameter $v$ to achieve an average TF rate of 93\% in the laboratory, the meeting room, and the corridor. Then, we exploit the training models to evaluate the performance in these three environments.

\begin{figure}[t]
	\center
	\includegraphics[width=2.0in]{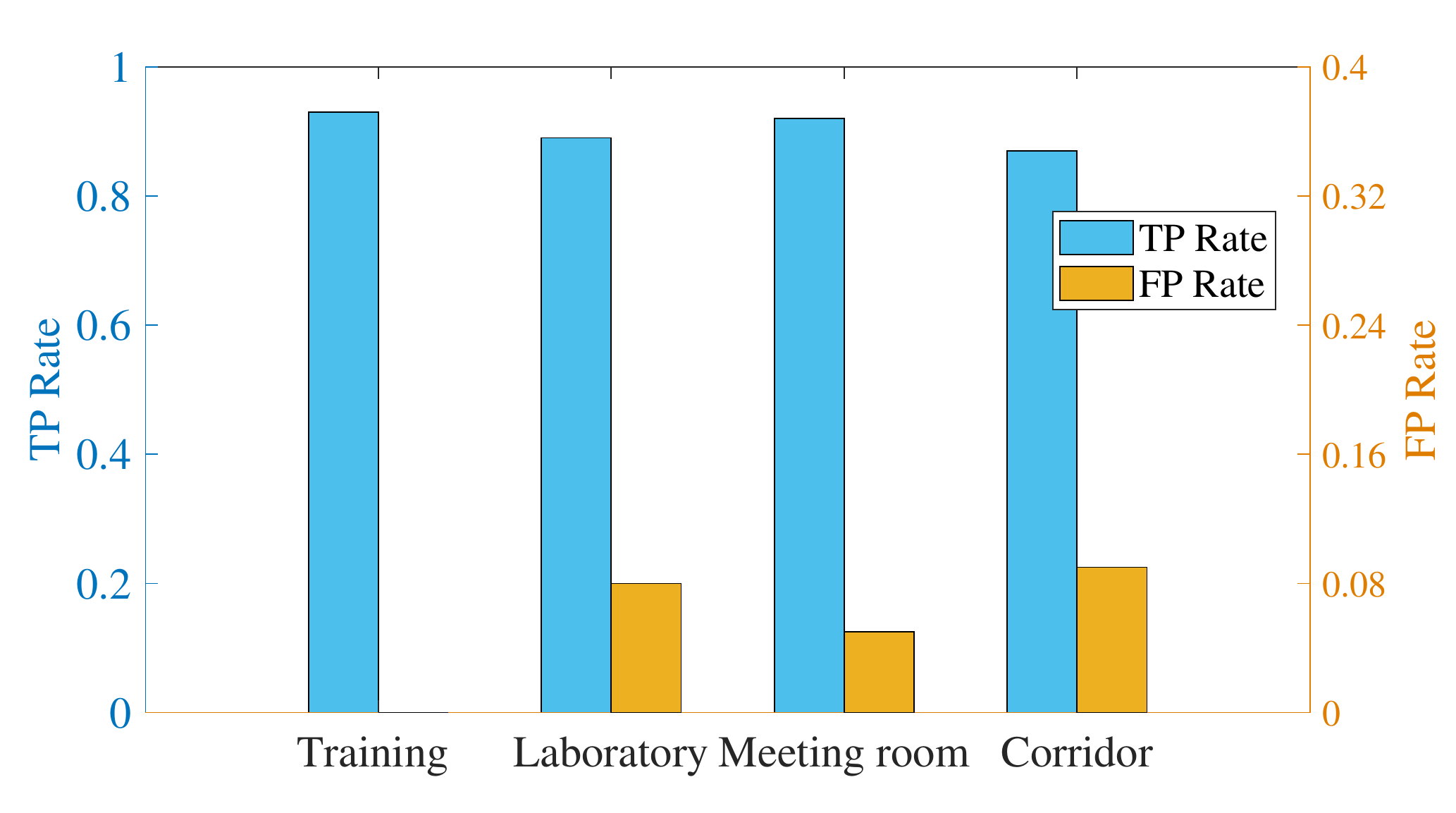} 
	\caption{The performance when we train without attacker samples.}
	\label{fig:no_negative}\vspace{-0.3cm}
\end{figure}

As described in Fig~\ref{fig:no_negative}, ShieldScatter can achieve an average TP rate of 89\%, and FP rate of 8\% in various environments, which are worse than the results by using the training model with some attacker samples. This is because the training model cannot well determine the boundary without using any attacker samples for constraint. In this case, we enable our system to repeat the process multiple times and authenticate the device with a voting scheme to improve the performance.

\subsection{ Experiment with Commodity Device}\label{sec:comm}
In this section, we also validate the performance by exploiting the commodity device. In particular, we install an off-the-shelf Intel 5300 WiFi card on an Intel Mini PC to act as the IoT device, which transmits the WiFi packets operating in 2.4GHz. Then, we attach three backscatter tags at a distance of 15cm from the receiver and control them to reflect the signals in turn. The positions of the transmitter and receiver remain the same as before. Finally, we place the attacker in various distances, locations, and directions from the IoT device to attack this system. At the same time, we extract the backscatter features from the collected WiFi packets and evaluate the performance. 

\begin{figure}[t]
	\center
	\includegraphics[width=2.0in]{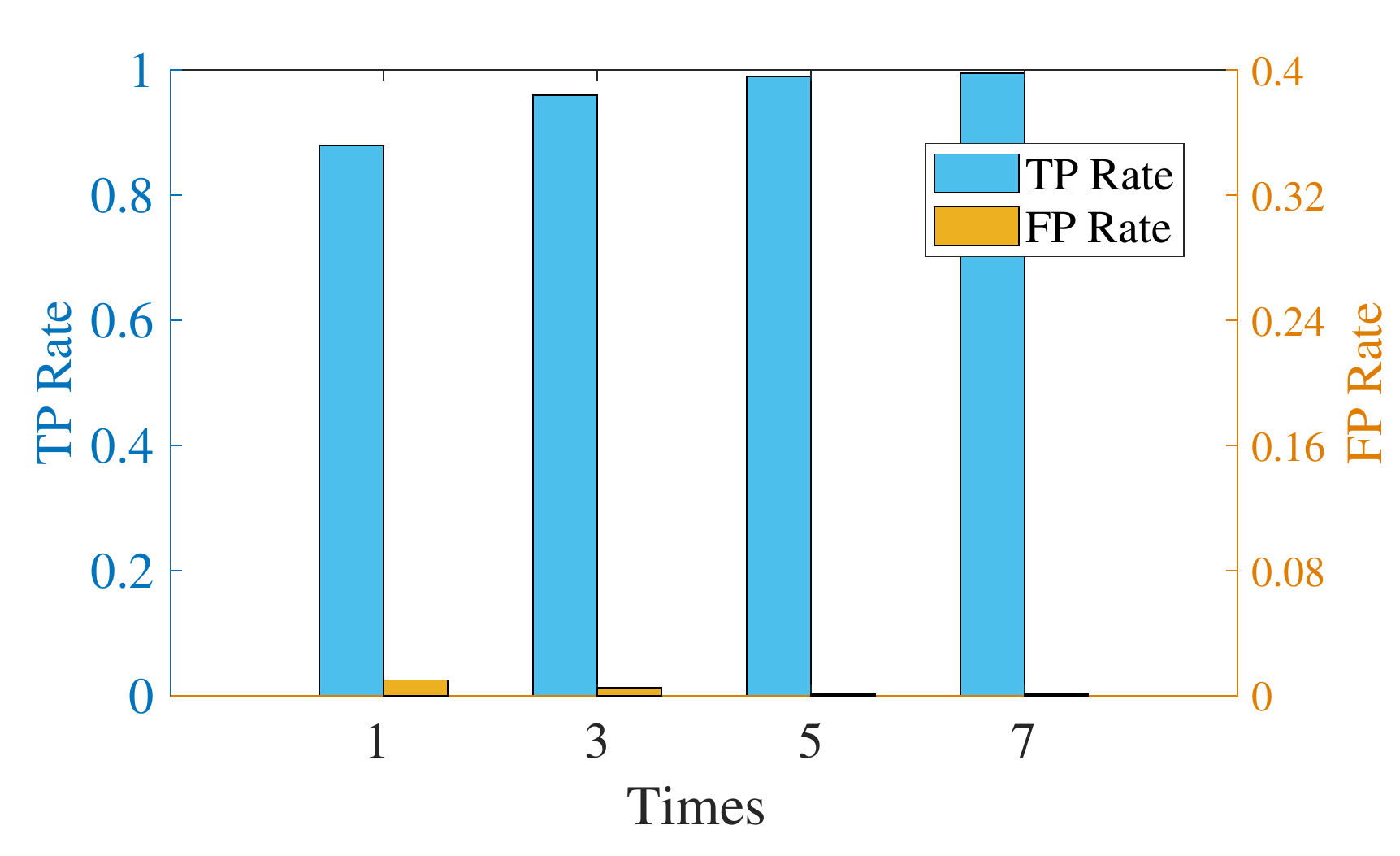} 
	\caption{The performance when we exploit commodity device for testing. }
	\label{fig:intel_com}\vspace{-0.3cm}
\end{figure}

The result is presented in Fig.~\ref{fig:intel_com}, where we can achieve the average TP rate around 88\%, which is lower than the results with USRP. The reason is that the imperfect design of the commodity WiFi card introduces a lot of noise into the signals, reducing the similarity of the legitimate signals. However, the hardware-caused noise also brings many differences between the legitimate IoT and attacker, and accordingly we can achieve a better FP rate of roughly 0\%. To improve the security, we also enable the system to authenticate the IoT device with a voting scheme, from which we can achieve the TP rate of 99\% and FP rate of 0\% with only five attempts. As a result, our system can also perform well with a commodity device.			

\section{Related work}\label{sec:related}
\textbf{ Backscatter communications.}
Backscatter communication has been considered to be a promising communication mechanism for the future. Backscatter originates from the RFID systems that exploit RFID readers to communicate with low-cost tags~\cite{wang2012efficient}. The difference between them is that backscatter can harvest ambient RF signals and enable two RF-powered devices to communicate by scattering and creating the path on the ambient signals~\cite{liu2013ambient}. Besides, in order to enable different RF-powered devices to communicate with each other, WiFi backscatter, FM backscatter and FS backscatter are proposed~\cite{kellogg2014wi,wang2017fm,zhang2016enabling}, making backscatter communication become applicable for current IoT devices. Recent efforts have designed backscatter tags that can be read with the commodity AP in a long range. For example, HitchHike~\cite{ zhang2016enabling} , FreeRider~\cite{ zhang2017freerider}, and Witag~\cite{ abedi2018witag } can achieve the communication ranges from 10-54m, which can provide pervasive connectivity service for high-density APs deployment. In our study, we have designed the backscatter tag that can reflect the signal at a distance larger than 30m to create significant multi-path propagation signatures and construct a unique profile for an IoT device.

\textbf{Channel similarity schemes.} Using wireless channel information for location distinction has been explored for many years~\cite{patwari2007robust, liu2010authenticating}. However, the work in~\cite{fang2014you} has proved that these channel comparison methods are not absolutely accurate. That is because an active attacker can precisely calculate the channel information and then break these approaches by emulating the multi-path signatures. The advantage of Shieldscatter is that this method makes it difficult for the attackers to estimate the real multi-path signatures created by the backscatter tags. Even though the attacker can emulate all the multi-path signatures or estimate the channel states from the legitimate IoT device to the receiver, it still cannot decide which multi-path signature is the true one each time while launching an attack. Thus, ShieldScatter is more reliable than the methods that simply compare the channel correlation.

\textbf{Physical-layer propagation signatures.}
In recent years, fine-grained physical-layer propagation signatures~\cite{wang2018securing,luo2018authenticating,huang2020lightweight} have been successfully used to secure wireless systems.  For example, SecureArray~\cite{ xiong2013securearray} secures the system by using AoA information to construct sensitive signatures. However, it needs more than 8 antennas and requires a complex calibration for the antennas each time when initiating the system, which is inappropriate to place such a large system in the smart home and is unacceptable for the user to repeat to calibrate this system. Besides, some researchers seek for other signatures, such as received signal strength (RSS) for user authentication~\cite{cai2011good,pierson2016wanda}. However, these studies need two or more antennas to construct the signatures, which would be not appropriate for smart home IoT devices that contain only one antenna. Different from these systems, ShieldScatter can secure the IoT devices without changing the hardware of the AP or the IoT to create an antenna array and it can still achieve high detecting accuracy with several low-cost backscatter tags for assistance.

\textbf{Localization.}
Accurate localization can help detect the signals and secure the IoT devices. SpotFi and ArrayTrack~\cite{kotaru2015spotfi,xiong2013arraytrack} can achieve high localization accuracy for the devices by combining AoA signatures with time of flight (ToF). WiTag~\cite{kotaru2017localizing} localizes the backscatter tags using commodity WiFi signals. However, these works need a large antenna array for help. It is unacceptable for some smart home devices.  Other localization systems, such as RFID, are also explored to localize the IoT devices~\cite{wang2013dude,wang2013rf}. All these RFID-based methods need a dedicated RFID reader to help communicate with the tags. In addition, there have been some explorations on localization depending on other signals. In ~\cite{liu2016guoguo,huang2015swadloon}, these works exploited acoustic signals for help. Besides, some other works~\cite{zhang2018visible,yang2019polarization} leverage visible light to localize the device. However, these methods may not be appropriate for low-power IoT devices due to the lack of hardware support.

\section{Conclusion}\label{sec:conclusion}			
We present ShieldScatter, a lightweight system to secure IoT device pairing and data transmission by intentionally creating multi-path signatures using several low-cost backscatter tags that are attached to an AP. ShieldScatter secures IoT devices without using a large antenna array or hardware modification to existing devices. We have evaluated the performance of ShieldScatter in both static and dynamic environments.  The experimental results show that even when the attacker is close to the legitimate device, ShieldScatter can mitigate 95\% of attack attempts while at the same time triggering false alarms on just 7\% of legitimate traffic.

\ifCLASSOPTIONcompsoc
\else
\fi


\ifCLASSOPTIONcaptionsoff
  \newpage
\fi

\balance
\bibliographystyle{IEEEtran}
\bibliography{sample-bibliography}

\begin{thebibliography}{10}
\providecommand{\url}[1]{#1}
\csname url@samestyle\endcsname
\providecommand{\newblock}{\relax}
\providecommand{\bibinfo}[2]{#2}
\providecommand{\BIBentrySTDinterwordspacing}{\spaceskip=0pt\relax}
\providecommand{\BIBentryALTinterwordstretchfactor}{4}
\providecommand{\BIBentryALTinterwordspacing}{\spaceskip=\fontdimen2\font plus
\BIBentryALTinterwordstretchfactor\fontdimen3\font minus
  \fontdimen4\font\relax}
\providecommand{\BIBforeignlanguage}[2]{{%
\expandafter\ifx\csname l@#1\endcsname\relax
\typeout{** WARNING: IEEEtran.bst: No hyphenation pattern has been}%
\typeout{** loaded for the language `#1'. Using the pattern for}%
\typeout{** the default language instead.}%
\else
\language=\csname l@#1\endcsname
\fi
#2}}
\providecommand{\BIBdecl}{\relax}
\BIBdecl

\bibitem{luo2018shieldscatter}
Z.~Luo, W.~Wang, J.~Qu, T.~Jiang, and Q.~Zhang, ``Shieldscatter: Improving iot
  security with backscatter assistance,'' in \emph{Proc. 16th ACM Conf.
  Embedded Netw. Sensor Syst.}, 2018, pp. 185--198.

\bibitem{8302842}
W.~{Wang}, L.~{Yang}, and Q.~{Zhang}, ``Resonance-based secure pairing for
  wearables,'' \emph{IEEE Trans. Mobile Comput.}, vol.~17, no.~11, pp.
  2607--2618, Nov 2018.

\bibitem{bertka2012802}
B.~Bertka, ``802.11 w security: Dos attacks and vulnerability controls,'' in
  \emph{Proc. of Infocom}, 2012.

\bibitem{tews2009practical}
E.~Tews and M.~Beck, ``Practical attacks against wep and wpa,'' in
  \emph{Proceedings of the second ACM conference on Wireless network security},
  2009, pp. 79--86.

\bibitem{mantas2011security}
G.~Mantas, D.~Lymberopoulos, and N.~Komninos, ``Security in smart home
  environment,'' in \emph{Wireless Technologies for Ambient Assisted Living and
  Healthcare: Systems and Applications}.\hskip 1em plus 0.5em minus 0.4em\relax
  IGI Global, 2011, pp. 170--191.

\bibitem{gehrmann2004manual}
C.~Gehrmann, C.~J. Mitchell, and K.~Nyberg, ``Manual authentication for
  wireless devices,'' \emph{RSA Cryptobytes}, vol.~7, no.~1, pp. 29--37, 2004.

\bibitem{dietrich2007financial}
S.~Dietrich and R.~Dhamija, ``Financial cryptography and data security,''
  \emph{Lecture Notes in Computer Science}, vol. 4886, 2007.

\bibitem{xiong2013securearray}
J.~Xiong and K.~Jamieson, ``Securearray: Improving wifi security with
  fine-grained physical-layer information,'' in \emph{Proc. 19th ACM Int. Conf.
  Mobile Comp. Netw.}, 2013, pp. 441--452.

\bibitem{jiang2013rejecting}
Z.~Jiang, J.~Zhao, X.-Y. Li, J.~Han, and W.~Xi, ``Rejecting the attack: Source
  authentication for wi-fi management frames using csi information,'' in
  \emph{Proc. IEEE INFOCOM}, 2013, pp. 2544--2552.

\bibitem{cai2011good}
L.~Cai, K.~Zeng, H.~Chen, and P.~Mohapatra, ``Good neighbor: Secure pairing of
  nearby wireless devices by multiple antennas,'' in \emph{Proc. Netw. Distrib.
  Syst. Security Symp.}, 2011.

\bibitem{xiao2009channel}
L.~Xiao, L.~J. Greenstein, N.~B. Mandayam, and W.~Trappe, ``Channel-based
  detection of sybil attacks in wireless networks,'' \emph{IEEE Trans. Inf.
  Forensics Security}, vol.~4, no.~3, pp. 492--503, 2009.

\bibitem{fang2014you}
S.~Fang, Y.~Liu, W.~Shen, and H.~Zhu, ``Where are you from?: confusing location
  distinction using virtual multipath camouflage,'' in \emph{Proc. 20th ACM
  Int. Conf. Mobile Comp. Netw.}, 2014, pp. 225--236.

\bibitem{liu2013ambient}
V.~Liu, A.~Parks, V.~Talla, S.~Gollakota, D.~Wetherall, and J.~R. Smith,
  ``Ambient backscatter: wireless communication out of thin air,'' in
  \emph{Proc. ACM SIGCOMM Conf.}, vol.~43, no.~4, 2013, pp. 39--50.

\bibitem{steele1999mobile}
R.~Steele and L.~Hanzo, \emph{Mobile Radio Communications: Second and Third
  Generation Cellular and WATM Systems: 2nd}.\hskip 1em plus 0.5em minus
  0.4em\relax IEEE Press-John Wiley, 1999.

\bibitem{wang2013dude}
J.~Wang and D.~Katabi, ``Dude, where's my card?: Rfid positioning that works
  with multipath and non-line of sight,'' in \emph{Proc. ACM SIGCOMM Conf.},
  vol.~43, no.~4, 2013, pp. 51--62.

\bibitem{salvador2007toward}
S.~Salvador and P.~Chan, ``Toward accurate dynamic time warping in linear time
  and space,'' \emph{Intelligent Data Analysis}, vol.~11, no.~5, pp. 561--580,
  2007.

\bibitem{scholkopf2001estimating}
B.~Sch{\"o}lkopf, J.~C. Platt, J.~Shawe-Taylor, A.~J. Smola, and R.~C.
  Williamson, ``Estimating the support of a high-dimensional distribution,''
  \emph{Neural computation}, vol.~13, no.~7, pp. 1443--1471, 2001.

\bibitem{eian2011modeling}
M.~Eian and S.~F. Mj{\o}lsnes, ``The modeling and comparison of wireless
  network denial of service attacks,'' in \emph{Proc. ACM SOSP workshop on
  networking, systems, and applications on mobile handhelds}.\hskip 1em plus
  0.5em minus 0.4em\relax ACM, 2011, p.~7.

\bibitem{tse2005fundamentals}
D.~Tse and P.~Viswanath, \emph{Fundamentals of wireless communication}.\hskip
  1em plus 0.5em minus 0.4em\relax Cambridge university press, 2005.

\bibitem{pierson2018poster}
T.~J. Pierson, T.~Peters, R.~Peterson, and D.~Kotz, ``Poster: Proximity
  detection with single-antenna iot devices,'' in \emph{Proc. 24th ACM Int.
  Conf. Mobile Comp. Netw.}, 2018, pp. 663--665.

\bibitem{wang2012efficient}
J.~Wang, H.~Hassanieh, D.~Katabi, and P.~Indyk, ``Efficient and reliable
  low-power backscatter networks,'' in \emph{Proc. ACM SIGCOMM Conf.}, 2012,
  pp. 61--72.

\bibitem{kellogg2014wi}
B.~Kellogg, A.~Parks, S.~Gollakota, J.~R. Smith, and D.~Wetherall, ``Wi-fi
  backscatter: Internet connectivity for rf-powered devices,'' in \emph{Proc.
  ACM SIGCOMM Conf.}, vol.~44, no.~4, 2014, pp. 607--618.

\bibitem{wang2017fm}
A.~Wang, V.~Iyer, V.~Talla, J.~R. Smith, and S.~Gollakota, ``Fm backscatter:
  Enabling connected cities and smart fabrics.'' in \emph{Proc. USENIX Conf.
  Netw. Syst. Des. Implementation}, 2017, pp. 243--258.

\bibitem{zhang2016enabling}
P.~Zhang, M.~Rostami, P.~Hu, and D.~Ganesan, ``Enabling practical backscatter
  communication for on-body sensors,'' in \emph{Proc. ACM SIGCOMM Conf.}, 2016,
  pp. 370--383.

\bibitem{zhang2017freerider}
P.~Zhang, C.~Josephson, D.~Bharadia, and S.~Katti, ``Freerider: Backscatter
  communication using commodity radios,'' in \emph{Proceedings of the 13th
  International Conference on emerging Networking EXperiments and
  Technologies}, 2017, pp. 389--401.

\bibitem{abedi2018witag}
A.~Abedi, M.~H. Mazaheri, O.~Abari, and T.~Brecht, ``Witag: Rethinking
  backscatter communication for wifi networks,'' in \emph{Proceedings of the
  17th ACM Workshop on Hot Topics in Networks}, 2018, pp. 148--154.

\bibitem{patwari2007robust}
N.~Patwari and S.~K. Kasera, ``Robust location distinction using temporal link
  signatures,'' in \emph{Proc. 13th ACM Int. Conf. Mobile Comp. Netw.}, 2007,
  pp. 111--122.

\bibitem{liu2010authenticating}
Y.~Liu, P.~Ning, and H.~Dai, ``Authenticating primary users' signals in
  cognitive radio networks via integrated cryptographic and wireless link
  signatures,'' in \emph{Proc. IEEE Security and Privacy (SP)}, 2010, pp.
  286--301.

\bibitem{wang2018securing}
W.~Wang, L.~Yang, Q.~Zhang, and T.~Jiang, ``Securing on-body iot devices by
  exploiting creeping wave propagation,'' \emph{IEEE J. Sel. Areas Commun.},
  vol.~36, no.~4, 2018.

\bibitem{luo2018authenticating}
Z.~Luo, W.~Wang, J.~Xiao, Q.~Huang, Q.~Zhang \emph{et~al.}, ``Authenticating
  on-body backscatter by exploiting propagation signatures,'' \emph{Proceedings
  of the ACM on Interactive, Mobile, Wearable and Ubiquitous Technologies},
  vol.~2, no.~3, p. 123, 2018.

\bibitem{huang2020lightweight}
Y.~Huang, W.~Wang, Y.~Wang, T.~Jiang, and Q.~Zhang, ``Lightweight
  sybil-resilient multi-robot networks by multipath manipulation,'' in
  \emph{IEEE INFOCOM 2020-IEEE Conference on Computer Communications}.\hskip
  1em plus 0.5em minus 0.4em\relax IEEE, 2020, pp. 2185--2193.

\bibitem{pierson2016wanda}
T.~J. Pierson, X.~Liang, R.~Peterson, and D.~Kotz, ``Wanda: securely
  introducing mobile devices,'' in \emph{Proc. IEEE INFOCOM}, 2016, pp. 1--9.

\bibitem{kotaru2015spotfi}
M.~Kotaru, K.~Joshi, D.~Bharadia, and S.~Katti, ``Spotfi: Decimeter level
  localization using wifi,'' in \emph{Proc. ACM SIGCOMM Conf.}, vol.~45, no.~4,
  2015, pp. 269--282.

\bibitem{xiong2013arraytrack}
J.~Xiong and K.~Jamieson, ``Arraytrack: A fine-grained indoor location
  system,'' in \emph{Proc. USENIX Conf. Netw. Syst. Des. Implementation}, 2013,
  pp. 71--84.

\bibitem{kotaru2017localizing}
M.~Kotaru, P.~Zhang, and S.~Katti, ``Localizing low-power backscatter tags
  using commodity wifi,'' in \emph{Proc. ACM CoNEXT}, 2017, pp. 251--262.

\bibitem{wang2013rf}
J.~Wang, F.~Adib, R.~Knepper, D.~Katabi, and D.~Rus, ``Rf-compass: Robot object
  manipulation using rfids,'' in \emph{Proc. 19th ACM Int. Conf. Mobile Comp.
  Netw.}, 2013, pp. 3--14.

\bibitem{liu2016guoguo}
K.~Liu, X.~Liu, and X.~Li, ``Guoguo: Enabling fine-grained smartphone
  localization via acoustic anchors,'' \emph{IEEE Trans. Mobile Comput.},
  vol.~15, no.~5, pp. 1144--1156, 2016.

\bibitem{huang2015swadloon}
W.~Huang, Y.~Xiong, X.-Y. Li, H.~Lin, X.~Mao, P.~Yang, Y.~Liu, and X.~Wang,
  ``Swadloon: Direction finding and indoor localization using acoustic signal
  by shaking smartphones,'' \emph{IEEE Trans. Mobile Comput.}, vol.~14, no.~10,
  pp. 2145--2157, 2015.

\bibitem{zhang2018visible}
C.~Zhang and X.~Zhang, ``Visible light localization using conventional light
  fixtures and smartphones,'' \emph{IEEE Trans. Mobile Comput.}, 2018.

\bibitem{yang2019polarization}
Z.~Yang, Z.~Wang, J.~Zhang, C.~Huang, and Q.~Zhang, ``Polarization-based
  visible light positioning,'' \emph{IEEE Trans. Mobile Comput.}, vol.~18,
  no.~3, pp. 715--727, 2019.

\end{thebibliography}
\end{document}